\documentclass[usenatbib]{mnras}
\usepackage{graphicx}
\usepackage{pdflscape}

\newcommand\msun{{\rm M_{\odot}}}

\def\go{
\mathrel{\raise.3ex\hbox{$>$}\mkern-14mu\lower0.6ex\hbox{$\sim$}}
}
\def\lo{
\mathrel{\raise.3ex\hbox{$<$}\mkern-14mu\lower0.6ex\hbox{$\sim$}}
}
\title[Swift observations of RS Oph]
{The 2021 outburst of the recurrent nova RS Ophiuchi observed in X-rays by the {\em Neil Gehrels Swift Observatory}: a comparative study}
\author[K.L. Page et al.]{K.L. Page$^{1}$, A.P. Beardmore$^{1}$, J.P. Osborne$^{1}$, U. Munari$^{2}$, J.-U. Ness$^{3}$, P.A. Evans$^{1}$, \newauthor M.F. Bode$^{4,5}$, M.J. Darnley$^{4}$, J.J. Drake$^{6}$, N.P.M. Kuin$^{7}$, T.J. O'Brien$^{8}$, M. Orio$^{9,10}$, \newauthor S.N. Shore$^{11,12}$, S. Starrfield$^{13}$ and C.E. Woodward$^{14}$.    \\
  $^{1}$ School of Physics \&  Astronomy, University of Leicester, LE1 7RH, UK\\
  $^{2}$ INAF National Institute of Astrophysics, 36012 Asiago, Italy\\
$^{3}$ European Space Agency (ESA), European Space Astronomy Centre (ESAC), Camino Bajo del Castillo s/n, E-28692 \\Villanueva de la Ca{\~ n}ada, Madrid, Spain\\  
  $^{4}$ Astrophysics Research Institute, Liverpool John Moores University, IC2 Liverpool Science Park, Liverpool L3 5RF, UK\\
  $^{5}$ Vice Chancellor's Office, Botswana International University of Science and Technology, Private Bag 16, Palapye, Botswana\\
  $^{6}$ Harvard-Smithsonian Center for Astrophysics, 60 Garden Street, Cambridge, MA 02138, USA\\
$^{7}$  Mullard Space Science Laboratory, University College London, Holmbury St. Mary, Dorking, Surrey RH5 6NT, UK\\
  $^{8}$ Jodrell Bank Centre for Astrophysics, Alan Turing Building, University of Manchester, Manchester, M13 9PL, UK\\
  $^{9}$   Department of Astronomy, University of Wisconsin-Madison, 475 N. Charter Street, Madison WI 53706, USA\\
  $^{10}$ INAF-Padova, vicolo Osservatorio, 5, I-35122 Padova, Italy\\
  $^{11}$ Dipartimento di Fisica `Enrico Fermi', Universit\`{a} di Pisa, I-56127 Pisa, Italy\\
$^{12}$ INFN-Sezione Pisa, largo B. Pontecorvo 3, I-56127 Pisa, Italy\\
  $^{13}$   Earth and Space Exploration, Arizona State University, P.O. Box 871404, Tempe, AZ 85287-1404, USA\\ 
$^{14}$ Minnesota Institute for Astrophysics, University of Minnesota, 116 Church Street SE, Minneapolis, MN 55455, USA\\
}
\date{Accepted XXX. Received YYY; in original form ZZZ}

\pubyear{2022}

\begin{document}
\label{firstpage}
\pagerange{\pageref{firstpage}--\pageref{lastpage}}

\maketitle

\begin{abstract}

On 2021 August 8, the recurrent nova RS Ophiuchi erupted again, after an interval of 15.5~yr. Regular monitoring by the {\em Neil Gehrels Swift Observatory} began promptly, on August 9.9 (0.37~day after the optical peak), and continued until the source passed behind the Sun at the start of November, 86 days later. Observations then restarted on day 197, once RS~Oph emerged from the Sun constraint.  This makes RS~Oph the first Galactic recurrent nova to have been monitored by {\em Swift} throughout two eruptions. Here we investigate the extensive X-ray datasets from 2006 and 2021, as well as the more limited data collected by {\em EXOSAT} in 1985. The hard X-rays arising from shock interactions between the nova ejecta and red giant wind are similar following the last two eruptions. In contrast, the early super-soft source (SSS) in 2021 was both less variable and significantly fainter than in 2006. However, 0.3--1~keV light-curves from 2021 reveal a 35~s quasi-periodic oscillation consistent in frequency with the 2006 data. The {\em Swift} X-ray spectra from 2021 are featureless, with the soft emission typically being well parametrized by a simple blackbody, while the 2006 spectra showed much stronger evidence for superimposed ionized absorption edges. Considering the data after day 60 following each eruption, during the supersoft phase the 2021 spectra are hotter, with smaller effective radii and lower wind absorption, leading to an apparently reduced bolometric luminosity. We explore possible explanations for the gross differences in observed SSS behaviour between the 2006 and 2021 outbursts.

\end{abstract}

\begin{keywords}
stars: individual: RS Oph -- novae, cataclysmic variables -- X-rays: stars 
\end{keywords}

\section{Introduction}
\label{intro}

Novae are the most energetic category of accreting white dwarfs (WDs).
They occur in interacting binary systems, within which mass transfer from the donor (secondary) star forms a layer of hydrogen on the WD (primary) surface. When enough material has been accreted, nuclear burning ignites at the base of the envelope, and, once sufficient pressure has built up, a thermonuclear runaway (TNR) begins \cite[see review articles in][]{bode08,woudt14}. The nova explosion expels material which initially obscures the WD surface from view. However, as the ejecta expand, they become optically thin, thus allowing any continuing nuclear burning on the WD to become visible. This emission ultimately peaks in the soft X-ray band, and is known as the Super-Soft Source (SSS) state \citep[e.g.,][]{krautter08}. The launch of the {\em Neil Gehrels Swift Observatory} \citep[{\em Swift} hereafter; ][]{geh04}, has led to this SSS phase being well monitored in many novae \citep[see][for recent reviews]{julo15, page20}, revealing previously unseen phenomena, such as high-amplitude flux variability and short timescale quasi-periodic modulations.

Classical novae (CNe) are those which have only been seen to erupt once. A small number of systems, however, have been detected in outburst multiple times, and these are known as recurrent novae (RNe). There are ten confirmed Galactic recurrents, of which the subject of this paper, RS Ophiuchi (RS~Oph hereafter), is one. It is believed that all novae should eventually recur, but most over periods of thousands of years (\citealt{yaron05} discuss the parameter space of nova outbursts). The Galactic RNe have recurrence timescales of $\sim$~10--100~yr (the upper bound being a selection effect dependent on historical records), while some RNe outside our own Galaxy have been found to recur as frequently as every year \cite[e.g.,][]{darn14}. Reviews of the recurrent novae in the Milky Way are provided by \cite{brad10} and \cite{darn21}.

The RS~Oph system comprises a massive WD and a red giant (RG) donor. The mass range estimate for the WD is 1.2--1.4~$\msun$ \citep[e.g.,][]{miko17}, and the orbital period, P$_{\rm orb}$, of the binary is 453.6~$\pm$~0.4~d \citep{brandi09}. Using the ephemeris in that paper, the eruption times in 2006 and 2021 are separated by (12.47~$\pm$~0.01)~P$_{\rm orb}$ -- that is, the orbital phase is close to 180$^{\circ}$ different between the two outbursts; the 1985 eruption occurred at approximately the same phase as in 2006. The RS~Oph WD and RG were in quadrature at each of these times (phases 0.26 and 0.72, respectively, where phase 0.0 is defined to be inferior conjunction of the RG). 

RS~Oph has previously passed through nova eruption cycles approximately every 15~yr, with a range spanning 9--27~yr \citep[see ][for a summary]{brad10}, having been first detected as a nova in 1898 \citep{pickering1905}. This short recurrence time suggests the system contains a massive WD, accreting at a substantial rate. Prior to 2021, the previous eruption occurred in 2006 Feb., and this was the first nova monitored in detail by {\em Swift}, leading to an exquisite dataset that was published by \cite{bode06} and \cite{julo11}. The detailed {\em Swift} results in 2006 revealed absorbed, hard X-rays early after the eruption, with the absorbing column decreasing and the temperature declining over the first few weeks. These spectra were explained as thermal emission from hot, shocked gas, as the nova ejecta interacted with the RG wind \citep{bode06}. On day 26, a new, very soft component was detected, signifying the beginning of the SSS phase. This soft emission was extremely variable in flux for almost three weeks, before settling at a consistently bright level until around day 60, when a monotonic decline set in.

Collimated bipolar structures (jets) were detected following the 2006 eruption \citep[e.g., ][]{obrien06, bode07, sok08, montez22}, highlighting the lack of spherical symmetry during the explosion \citep[see also models by][]{orlando09,walder08}. No eclipses are seen in RS~Oph, placing limits on the angle of inclination, with estimates ranging from 30--52$^{o}$ \citep{dobrzycka94, brandi09, valerio12}.

RS~Oph was reported to have erupted again on 2021 Aug. 08, as described in AAVSO\footnote{American Association of Variable Star Observers} Alert Notice 752\footnote{https://www.aavso.org/aavso-alert-notice-752}. This prompted a flurry of observations across the electromagnetic spectrum, from very high energy TeV/GeV $\gamma$-rays detected by H.E.S.S.\footnote{The High Energy Stereoscopic System} \citep{wagner21a, wagner21b,hess} and MAGIC\footnote{Major Atmospheric Gamma Imaging Cherenkov Telescopes} \citep{magic}, and MeV $\gamma$-rays by {\em Fermi}-LAT\footnote{Large Area Telescope} \citep{cheung21a, cheung21b, cheung22}; to X-rays seen by {\em MAXI}\footnote{Monitor of All-sky X-ray Image} \citep{maxi}, {\em INTEGRAL}\footnote{INTErnational Gamma-Ray Astrophysics Laboratory} \citep{integral}, {\em NICER}\footnote{Neutron Star Interior Composition Explorer Mission} \citep{nicer1, nicer2, nicernustar, nicer3, nicer4}, {\em NuSTAR}\footnote{Nuclear Spectroscopic Telescope Array} \citep{nicernustar}, {\em AstroSat} \citep{astrosat}, {\em Chandra} and {\em XMM-Newton} \citep{chandra}, as well as {\em Swift} \citep{page21, page21a, page21b}; to optical \citep{opt1, opt2, opt3, opt4, opt5, opt6, opt7, opt8, opt9, opt10, opt11, opt12, opt13}; infra-red \citep{ir1, ir2}; and radio \citep{radio1, radio2, radio3}. A neutrino search was performed by IceCube, though only upper limits could be placed \citep{icecube}.

In this paper we present the {\em Swift} X-ray observations of the 2021 outburst of RS~Oph, and compare them to the data obtained following the 2006 eruption, finding significant differences over the soft X-ray band, while the harder X-rays are much more similar. We focus mainly on the SSS emission, while \cite{cheung22} present a discussion of the higher-energy shock emission alongside the LAT results.

Despite the well-known close similarity between the optical light-curves of individual RNe \citep[e.g.,][]{brad10}, the SSS phase can sometimes show wider variations from one outburst to the next. For example, the yearly recurrent M31N 2008-12a showed very similar (optically identical) multi-wavelength results from 2008--2021, with the exception of the 2016 eruption, which was notably different in X-rays \citep[][Healy et al. in prep.]{henze18}. 
Nova LMC 1968 has also been observed by {\em Swift} through two eruptions (2016 and 2020), in both cases showing very similar X-ray (and optical) evolution \citep{kuin20, schwarz20}. RS~Oph is the first Galactic recurrent for which this comparison has been possible\footnote{We note that it appears that the latest eruption of U~Sco might have occurred during solar conjunction in 2017, and was thus entirely missed (B. Schaefer, priv. comm.), otherwise this event would also have been observed by {\em Swift}.}; it is also one of the brightest RNe in the X-ray band, thus allowing a detailed examination.

Throughout this paper, errors are given at the 1$\sigma$ confidence level,
 unless otherwise stated. Spectra were binned such that they have a
 minimum of 1 count~bin$^{-1}$ to facilitate Cash statistic
 \citep{cash79} fitting within {\sc xspec} \citep{arn96}, and the
 abundances from \cite{wilms00}, together with the photoelectric
 absorption cross-sections from \cite{vern96}, have been assumed when
 using the T{\" u}bingen-Boulder Interstellar Medium ({\sc xspec/tbabs}) absorption model.

 For ease of comparison with the previous publications for the {\em Swift} 2006 data, the time origin of the light-curves presented here, T0, has been taken as the time at which the optical emission peaked; this is estimated to be 2021 Aug. 09.542 UT (MJD 59435.542) from the AAVSO light-curve\footnote{For the LAT data analysis, \cite{cheung22} have assigned T0 to be the start time of the eruption from \cite{munari21}, 1.04 day before our optical peak time.}.
For 2006, T0 (again, the time of optical maximum) is taken to be 2006 Feb. 12.83 UT \citep{hirosawa06}. We also assume a distance of 1.6~kpc \citep{bode87}. The parallax determined by {\em Gaia} corresponds to a greater distance of 2.4$^{+0.3}_{-0.2}$~kpc \cite{gaia21}, but \cite{brad18} notes that these parallaxes are not yet reliable for systems with long-period binary orbits, and can lead to either under- or over-estimates of the distance. See \cite{montez22,magic} for recent discussions on the distance to RS~Oph.

\section{Observations}
\label{obs}

Following a Target of Opportunity request promptly after the announcement of the new eruption, {\em Swift} observations began on 2021 Aug 9.9, only 0.37 day after the optical peak \citep[1.41 day after the probable start time of the nova event;][]{munari21}. Initially, observations were performed on a daily basis; from the start of September, the cadence was increased to every 12~hr, then to every 8~hr between Sep. 15 and Oct. 1. Daily observations then resumed until RS~Oph entered the {\em Swift} solar observing constraint on Nov. 5. In addition, on Sep. 12 observations were taken every {\em Swift} orbit (approximately every 1.5~hr); this date was chosen as a time corresponding to the number of days ($\sim$~33) after the 2006 eruption when high-amplitude soft flux variability was seen. These observations were all obtained using the X-ray Telescope \cite[XRT; ][]{xrt} Windowed Timing (WT) mode, since the count rate was consistently above 1~count~s$^{-1}$. There were, in addition, a number of separate observations performed in Photon Counting (PC) mode between Sep. 24 and Oct. 08; these were taken to investigate possible extended emission around the source. Given the extreme level of pile-up in these PC observations, the data are not considered in this paper.

\begin{figure}
\begin{center}
\includegraphics[clip, width=5.5cm, angle=-90]{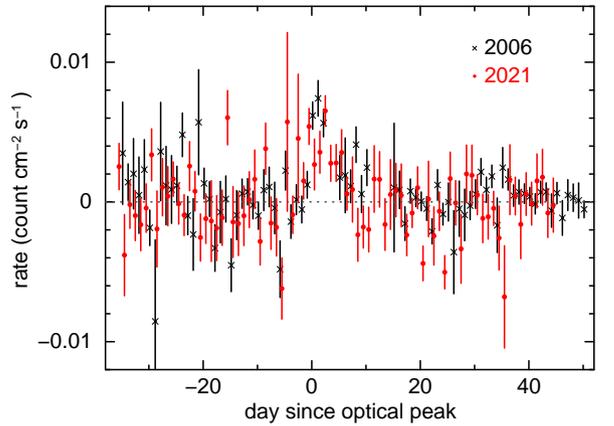}
\caption{BAT data from the Transient Monitor, showing a faint, but clear, detection of RS~Oph over 15--50~keV in both 2006 (black crosses) and 2021 (red circles). The abscissa shows days since the peak in the optical light-curve for each eruption. The horizontal dotted line marks the zero count rate level.}
\label{bat}
\end{center}
\end{figure}

Photometric and grism data were also collected using the {\em Swift} UV/Optical telescope \cite[UVOT; ][]{uvot}. These results will be presented in a separate publication. In addition, the Burst Alert Telescope \cite[BAT; ][]{bat} Transient Monitor\footnote{https://swift.gsfc.nasa.gov/results/transients/RSOph/} \citep{krimm13} showed an apparent weak detection of RS~Oph over 15--50~keV for a few days (Fig.~\ref{bat}) around the time of the eruption \citep[as was the case in 2006; ][]{bode06}.

\begin{figure}
\begin{center}
\includegraphics[width=8cm]{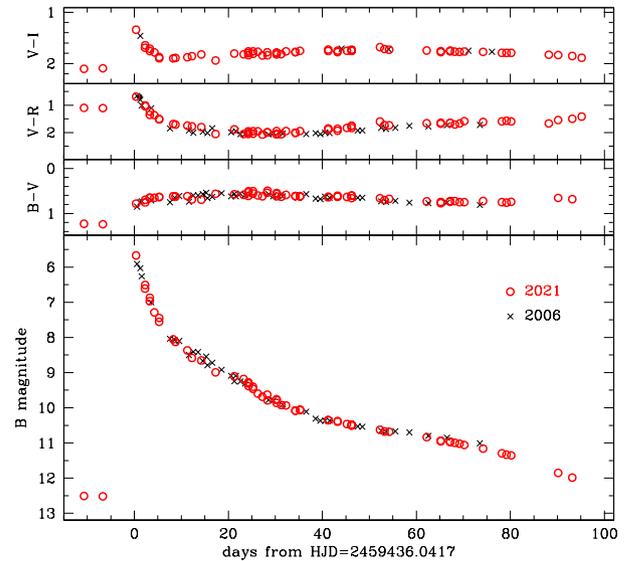}
\caption{$B$-band photometry and colour diagrams from the ANS Collaboration, from both 2021 and 2006. The two outbursts are almost identical at optical wavelengths.}
\label{opt}
\end{center}
\end{figure}

Fig.~\ref{opt} shows the optical data collected by the Asiago Novae and Symbiotic stars (ANS) Collaboration during both 2021 and 2006 \citep[see][for a summary of the 2006 results]{munari07}. The light-curves and colour evolution from the two eruptions appear almost identical. It can be seen that, at the time RS~Oph entered solar conjunction in 2021, on day 86, the system had not yet returned to the pre-eruption brightness, still being around half a magnitude brighter in $B$.

The flux of the optical [Fe\,{\sc x}] 6375~\AA\ coronal line was measured from the RS~Oph spectral atlas by \cite{munari21,munari22}, which provides absolute fluxed spectra (primarily Echelle) obtained at nearly daily cadence. These results are discussed in Section~\ref{comp}.

The {\em Swift} XRT data were analysed with HEASoft 6.29 and the most up-to-date calibration files available at the time (release date of 2021 Sep. 23). Only grade 0 (single pixel) events were utilised, in order to help minimise optical loading\footnote{https://www.swift.ac.uk/analysis/xrt/optical\_loading.php}. We note that the 2006 data presented in \cite{julo11} used WT grades 0--2 (at that time our understanding of the best way to mitigate optical loading was less complete than it is today). The 2006 light-curves and spectra presented here have also been reprocessed with the latest software and only grades 0 considered; the results are not materially different from \cite{julo11}, however.

\section{X-ray light-curve}

Light-curves for both the 2006 and 2021 datasets were generated using the online XRT product generator\footnote{https://www.swift.ac.uk/user\_objects/} \citep{evans07, evans09}. Looking at the shape of the spectra, 1~keV was chosen as a suitable cut between the SSS component and the harder, shock-related emission. Fig.~\ref{lc} shows the light-curves over the soft (0.3--1~keV) and hard (1--10~keV) bands, together with the hardness ratios (defined as the ratio of the counts in the 1-10 keV and 0.3-1 keV bands) for the two eruptions. There are clearly some similarities between the datasets, particularly at earlier times before the start of the SSS phase, but, overall, the evolution of the soft-band light-curve is markedly different, with the bulk of the 2021 SSS emission rising later, peaking at a lower level and then starting to fade earlier, before rebrightening briefly, compared with 2006.

\subsection{0.3--1~keV}
\label{soft}

Basing our expectations of the light-curve evolution on the results from 2006, and in order to put stringent limits on the start of the SSS in 2021, observations with at least a daily cadence were obtained from the very beginning of our monitoring campaign. The first hint of spectral softening appeared on day 20.6 \citep{page21a}, with no sign of enhanced low-energy emission during the previous observation on day 19.5. This `bump' in the spectrum stayed small until day 26.3, when the soft component increased significantly \citep{page21b}. In 2006, (weak) soft emission was first detected on day 26.0, with an obvious increase in counts below 1~keV on day 29.0 \citep{julo11}. Before this, the previous observation occurred eight days earlier, on day 18.2. Therefore, although the 2021 SSS was noted a few days earlier than in 2006, this may simply be due to the denser sampling of the more recent eruption.

\begin{figure*}
\begin{center}
  \includegraphics[clip, width=10cm, angle=-90]{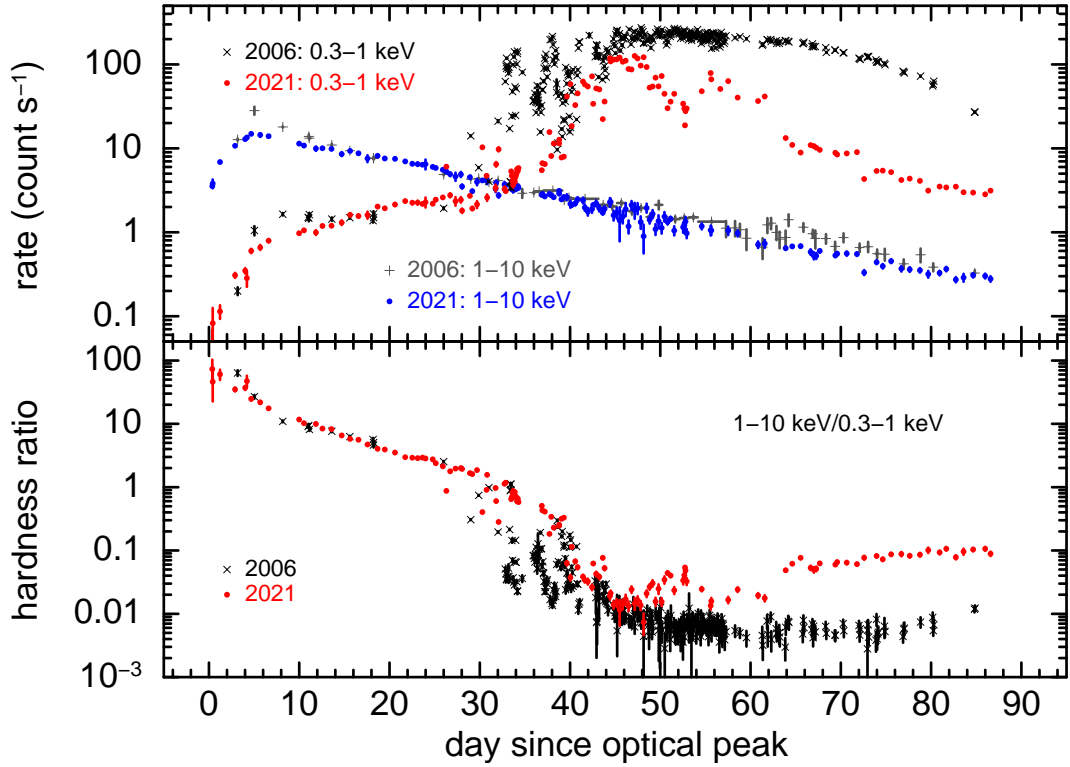}
\caption{Top: A comparison of the soft (0.3-1 keV: 2006 -- black crosses; 2021 -- red circles) and hard (1-10 keV: 2006 -- grey pluses, 2021 -- blue circles) light-curves following the 2006 and 2021 eruptions. Bottom: The hardness ratios following the 2006 (black crosses) and 2021 (red circles) eruptions.}
\label{lc}
\end{center}
\end{figure*}

\begin{figure}
\begin{center}
\includegraphics[clip, width=5.5cm, angle=-90]{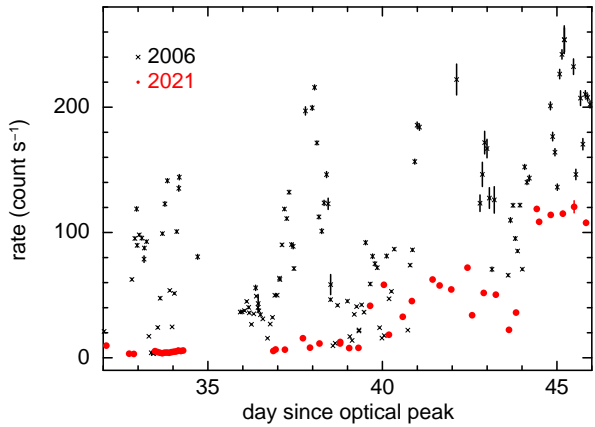}
\caption{A comparison of the 2006 and 2021 0.3--1~keV light-curves during the early supersoft phase. The 2006 dataset reaches a higher count rate, and shows larger amplitude changes, though both datasets are variable. The peaks and troughs of the variability occur at different times in 2006 and 2021.}
\label{sss}
\end{center}
\end{figure}

The onset of the SSS in 2006 was, however, much more dramatic than in 2021. Fig.~\ref{sss} zooms in on the soft light-curve between days 32 and 46, during which time the 2006 data showed high-amplitude flux variability, with changes in X-ray count-rate of more than an order of magnitude in $\sim$~12~hr \citep[see also ][]{julo11}. While the 2021 observations did show variability, Fig.~\ref{sss} shows this was at a significantly lower level, with a peak count rate less than half that measured in 2006. It is clear that the timing of peaks and troughs in 2006 and 2021 is not the same. This is expected if the flux variability is caused by absorption in random, clumpy ejecta, which was discussed as a possibility by \cite{julo11}.

The other big difference, as shown in Fig.~\ref{lc}, is that the 2021 soft X-ray emission showed a definite fading trend between days $\sim$~48--53, decreasing from around 100~count~s$^{-1}$ to $\sim$~20~count~s$^{-1}$, before rebrightening again over a few days. In 2006, the X-rays stayed consistently bright between days 46--58, after which time a steady decline set in \citep{julo11}. As RS~Oph entered the solar observing constraint in 2021 (day 87), the 0.3--10~keV XRT count rate was $\sim$~3.4~count~s$^{-1}$, while the count rate on the corresponding day in 2006 was $\sim$~20~count~s$^{-1}$.

The apparent rate of decline of the 2021 soft X-rays was also noticeably slower than the 2006 data at this late time: approximating the decay between days 75--85 with a power-law\footnote{Using a power-law is simply a numerical convenience, without any physical implication.} of t$^{-\alpha}$, the 2021 data can be fitted with $\alpha$~$\sim$~4.5, while the 2006 light-curve follows $\alpha$~$\sim$~11.5. If instead we consider all the data after day 62 (after which point both light-curves show a monotonic fading), the 2021 data decay can be approximated with a single power-law of $\alpha$~$\sim$~5.1 (although the data show considerable scatter), while the 2006 light-curve shows a break in its fading, with $\alpha$~$\sim$~3.6 until day 76, then steepening to $\sim$~11.5.

By comparison of the light-curves, we estimate the integrated number of observed soft (0.3--1~keV) X-ray counts between days 30--86 in 2006 was about 4--5 times larger than in 2021.

\subsubsection{QPO}
\label{sec:qpo}

\begin{figure}
\begin{center}
\includegraphics[clip, width=8cm]{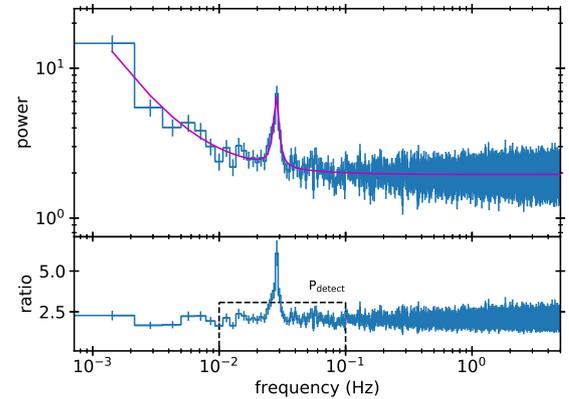}
\caption{The upper panel shows the average of 59 periodograms (each spanning
700~s) from the 2021 WT data over days 36.87 -- 61.56 (the main SSS
interval), fitted with the model described in the text (magenta line). The
bottom panel shows the ratio obtained between the average and model
when the Lorentz (QPO) term is removed. The dashed line shows the 99~per~cent detection level over the frequency range 0.01 -- 0.10~Hz (i.e., periods of 10 -- 100~s).}
\label{qpo}
\end{center}
\end{figure}

The 2006 eruption of RS~Oph revealed a strong $\sim$~35~s
quasi-periodic oscillation (QPO) during the SSS phase
\citep{andy08, julo11}. Short period oscillations were subsequently also
identified in other bright novae observed by {\em Swift}, with
timescales of up to $\sim$~100~s (e.g., KT~Eri -- \citealt{kteri}, V339~Del -- \citealt{v339del} and V5668~Sgr -- \citealt{v5668sgr}), and were confirmed in {\em XMM-Newton} and {\em Chandra} observations
\citep{ness15}.

In order to search for similar oscillations from the current outburst,
we extracted light curves from the WT data at 0.1~s binning in the 0.3--1~keV
energy band, and applied a standard Fast Fourier
Transform algorithm to calculate periodograms from individual
snapshots (that is, continuous on-target pointings) and their average \citep[e.g.][]{leahy83, vanderklis88, vaughan03}. The
periodograms were normalised following Leahy et al., so have
predictable Poisson noise properties allowing the significance of
potential signals to be evaluated. A maximum likelihood periodogram
fitting technique was then applied \cite[e.g.,][]{barret12} to
identify the QPO parameters. Given the mean duration of the light
curves from each snapshot was $\sim$~700~s, and ranged from 380--1250~s,
the periodograms were computed over 700~s continuous time bins and
padded to this duration with the mean snapshot count rate if they were
shorter.

The average periodogram from days 36.87--61.56 covering the SSS
bright phase is shown in Fig.~\ref{qpo}. The 35~s modulation is clearly
detected once more at greater than 99 per~cent confidence. By fitting
the periodogram with a model consisting of a powerlaw (for the low
frequency variations), a constant (for Poisson noise) and a Lorentzian
(for the QPO), a best fit oscillation frequency of 28.31~$\pm$~0.21~mHz was obtained, corresponding to a period of 35.33~$\pm$~0.26~s, with
a fractional rms of 1.94~$\pm$~0.2~per~cent.  The best fit width of
the Lorentzian corresponds to an average coherence of 7.7~$\pm$~1.8
cycles. The QPO was not detected before or after days 36.87--61.56. Examination of the individual periodograms shows the QPO was detected in
approximately 20~per~cent of this interval. The maximum amplitude seen
in any individual periodogram was 5~per~cent. {\em NICER} also reported the detection of a $\sim$~35~s oscillation \citep{nicer3}.

During the 2006 outburst, the XRT measured an average best fit period
of 35.04~$\pm$~0.09~s, a fractional rms of 2.43~$\pm$~0.20, and coherence
of 20.9~$\pm$~2.4 cycles (Beardmore et al. in prep. will provide a full analysis of the 2006 data). Thus, while the
QPO was not so strongly detected in 2021 (probably due in part to the
shorter snapshots of data obtained, as well as the source being
fainter), the periods are entirely consistent.

\subsection{1--10~keV}

In comparison with the large differences over the 0.3--1~keV band between the two eruptions, the 1--10~keV X-rays were much more consistent, in terms of both brightness and evolution, as Fig.~\ref{lc} shows. The 2021 {\em Swift} observations began almost three days earlier than in 2006, giving us a firmer handle on the initial rise of the X-rays. The harder emission was seen to increase in strength until around day 5 following both eruptions, though the 2021 peak 1--10~keV count rate is only about half that seen in 2006. (The higher cadence of observations in 2021 makes the peak time better defined.) The 2006 data then faded more rapidly until around day 15--20 (following an approximate power-law decay of $\alpha$~$\sim$~1, compared with 0.5 for the 2021 data), at which point both the count rate and rate of decline became very similar to what we see in 2021 ($\alpha$~$\sim$~2). There is an interval between days 60--70 where the 2006 data showed a slight excess compared with 2021 but, overall, the 1--10~keV light-curves are rather similar between the two nova events, especially in comparison with the differences in the soft emission.

\section{X-ray Spectrum}
\label{sec:spec}

\begin{figure*}
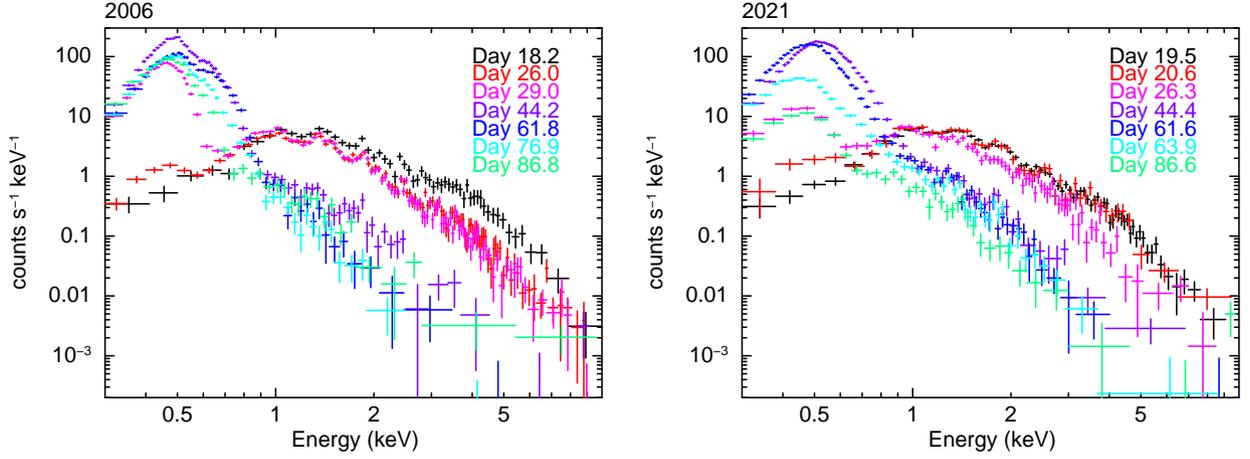

  \begin{center} 
    \includegraphics[clip, width=6cm, angle=-90]{RSOph_spec-comp_2006.ps}
    \includegraphics[clip, width=6cm, angle=-90]{RSOph_spec-comp_2021.ps}   
    \caption{Evolution of the X-ray spectra through the 2006 (left) and 2021 (right) eruptions. Left: 2006 spectra. The black spectrum shows the last date when there was no detectable SSS emission, on day 18.2, while the red shows the first where a soft excess was seen, on day 26.0, eight days later. The next observation occurred on day 29, and showed a strong soft component. Days 44.2, 61.8 and 86.8 are shown for direct comparison with the 2021 results on those days (see right-hand plot and description below). Since there was no sudden drop in observed count rate between days 61.6 and 63.9 in 2006, the spectrum from day 76.9 is included instead. Right: 2021 spectra. The black spectrum again shows the last date when there was no detectable SSS emission, on day 19.5; the next observation, in this case taken only 27 hours later, showed a slight increase in counts below $\sim$~0.6~keV. The slightly-enhanced soft emission stayed around this level until day 26.3, when an obvious soft component appeared. The X-ray count rate reached a peak around day 44.4, after which there was a decline followed by a rebrightening. Due to the proximity of RS Oph to the Moon, the source could not be observed between days 61.6 and 63.9, during which time the count rate dropped from $\sim$~41.5~count~s$^{-1}$ to $\sim$~14~count~s$^{-1}$ (blue and cyan spectra). The blue/green spectrum corresponds to the last day RS~Oph could be observed before entering the Sun constraint.}
\label{spec}
\end{center}
\end{figure*}

\begin{figure}
\begin{center}
  \includegraphics[clip, width=5.5cm, angle=-90]{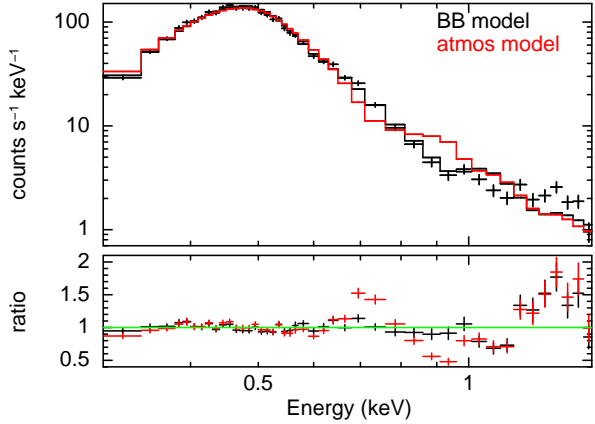}
\caption{Comparison of the BB (with no absorption edges; kT~=~43~$\pm$~1~eV) and atmosphere fits (using Rauch grid 003; kT~=~70~$\pm$~1~eV) to the SSS emission on day 43.8 of the 2021 eruption. The bottom panel plots the ratio of the data to both models; this clearly shows the fit deficiency using the Rauch grid 003 atmosphere model in the interval 0.7--1~keV}
\label{bb-atmos}
\end{center}
\end{figure}

Fig.~\ref{spec} shows samples of X-ray spectra obtained during the 2006 and 2021 eruptions.
The 2021 spectra show a very smooth, featureless super-soft component, which can be simply modelled by a blackbody (BB) component -- occasionally improved by the inclusion of absorption edge(s), though often well-modelled without. While \cite{julo11} parametrized the 2006 data by use of an atmosphere component \citep[the T{\" u}bingen non-local-thermodynamic-equilibrium model atmospheres;][]{rauch03}, a BB+edges model also provides an acceptable fit to these 2006 data -- in most cases, statistically better. In addition, the model atmospheres seemed unstable within {\sc xspec}, with the fitted temperatures appearing almost quantised at certain values, despite small step sizes for fitting, and the minimisation routine implemented.
Given that high-resolution observations of novae have shown that the currently available atmosphere models do not account for all the complexities seen \citep{ness19}, we have chosen to use the BB option for a direct comparison between the two eruptions. Fig.~\ref{bb-atmos} demonstrates the better fit results obtained using a pure BB compared to the T{\" u}bingen grid 003 model atmosphere \cite[as used in ][]{julo11}, particularly between 0.7--1~keV. The best fits for these two models applied to this example spectrum are kT$_{\rm BB}$~=~43~$\pm$~1~eV, R$_{\rm BB}$~=~(3.9$^{+0.6}_{-1.8}$)~$\times$~10$^{10}$~cm, N$_{\rm H,BB}$~=~(2.6$^{+0.2}_{-0.3}$)~$\times$~10$^{21}$~cm$^{-2}$, and kT$_{\rm atmos}$~= 70~$\pm$~1~eV, R$_{\rm atmos}$~=~(1.0~$\pm$~0.3)~$\times$~10$^{10}$~cm, N$_{\rm H, atmos}$~=~(2.9~$\pm$~0.2)~$\times$~10$^{21}$~cm$^{-2}$, respectively.

\begin{figure}
\begin{center}
  \includegraphics[clip, width=5.5cm, angle=-90]{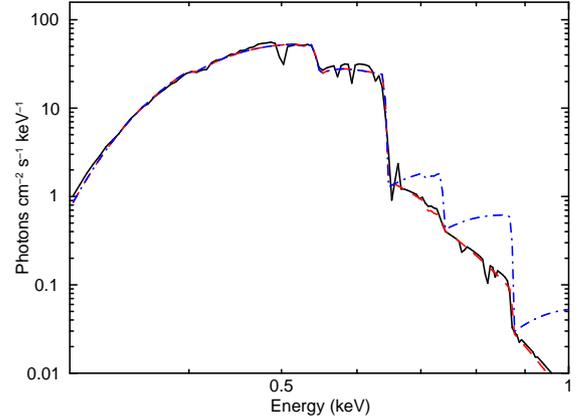}
  \caption{Comparison of the Rauch atmosphere grid (solid black line), the BB model including standard {\sc xspec} edges following E$^{-3}$ dependence (blue dot-dashed lines; this is not an actual fit to the data, but aligned to show the different edge shapes clearly) and the BB model including edges with an E$^{+1}$ recovery (red dashed lines). In the model used for this example, only additional edges at 0.635, 0.739 and 0.871~keV were significant; the dip close to 0.55~keV is simply due to the interstellar absorption modelled by {\sc tbabs}, so the shape does not vary between the models.}
\label{edges}
\end{center}
\end{figure}

Looking in detail at the appropriate temperature Rauch atmosphere models, we determined that
the strongest edges were those associated with He- and H-like N at
0.552 and 0.667~keV, and He- and H-like O at 0.739 and 0.871~keV. However, the
energy dependence of the edge profiles in these grid model is very different from
those implemented by the standard edge model in {\sc xspec}, which
assumes an E$^{-3}$ cross-section above the edge. From the atmosphere models, we
find a flatter dependence of E$^{+1}$ is more appropriate at the
temperatures of interest for RS~Oph, so a local {\sc xspec} model was defined to implement this different edge shape\footnote{Our cross-section energy dependance changes both the edge depth and the post-edge spectrum compared to the standard E$^{-3}$ profile; while trends
within our fits can be informative, direct comparisons with model fits using standard edges will not be meaningful.}. The differences are shown in Fig.~\ref{edges}. Additionally, the atmosphere
models have absorption lines which may overlap/interact with the edges, one of
which (H-like O at 0.654~keV) shifts the 0.667~keV edge
energy to 0.635~keV at XRT CCD level spectral resolution.

All four of these edges were implemented in our model (using the E$^{+1}$ shape). The optical depths of the edges accounting for He-like and H-like N (0.552 and 0.635~keV) and He-like O (0.739~keV) were allowed to vary over a range of $\tau$~=~0--5. The optical depth of the fourth edge (H-like O at 0.871~keV) included in the fit was fixed at 2; while this edge often improved the residuals, the overlap of the BB and cooler optically-thin component (see below) made constraining the depth difficult.

The high-resolution grating spectra obtained following the 2006 eruption show both absorption and emission lines superimposed on the continuum \citep{ness07a,ness09}, as do the 2021 spectra (Ness et al. in prep.). Such lines are not required when modelling the lower-resolution XRT spectra, though unmodelled emission lines at low energies could affect the fitted absorbing column. A quick comparison between the 2021 {\em XMM-Newton} RGS\footnote{Reflection Grating Spectrometer.} spectra and contemporaneous XRT spectra gives no great cause for concern at these low energies, however. Indeed, the fit parameters from the {\em Swift} data closest in time to the {\em XMM-Newton} observation on day 37 of the 2021 eruption, applied directly to those RGS spectra shows that they are a reasonable approximation to the underlying continuum, with emission lines superimposed. Similarly, a comparison between contemporaneous XRT and {\em NICER} spectra (Orio et al. in prep.) shows general agreement. 

We note that the resolution of the XRT WT mode at 0.5~keV is currently 115~eV (full width at half maximum -- FWHM), compared with 82~eV in 2006\footnote{The FWHM of 63~eV given in \cite{julo11} was measured using an older response function (RMF). Since that work was published, the calibration was
updated with a slightly broader response.}. 
Sharp spectral features will therefore not be resolved in the {\em Swift} data.

\begin{figure*}
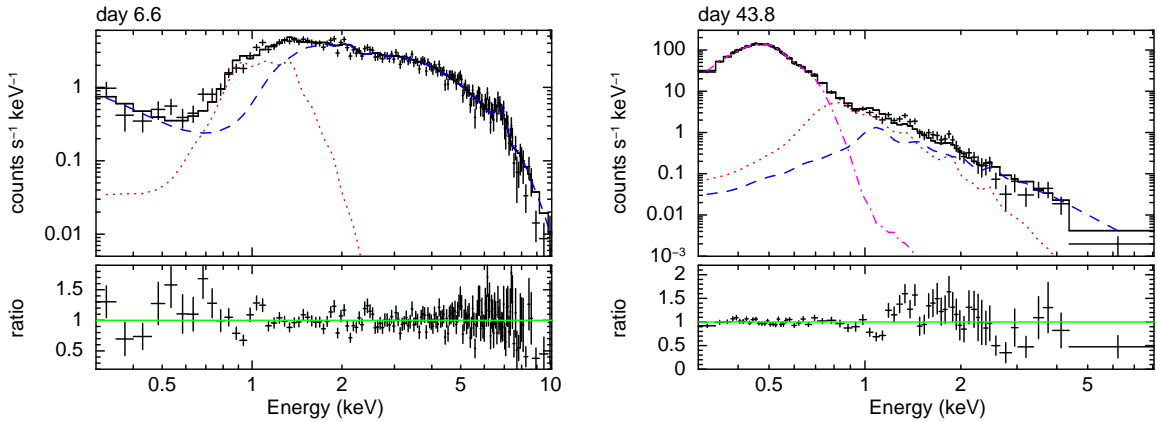

\begin{center}
  \includegraphics[clip, width=5.5cm, angle=-90]{day6.6_2apec.ps}  
  \includegraphics[clip, width=5.5cm, angle=-90]{day43.8_bbkedges2apec.ps}
\caption{Example spectra demonstrating how well the models fit the 2021 data, before and during the SSS phase. Left panel: Spectrum from day 6.6 after the eruption, at the time the 1--10~keV emission peaked, fitted with two optically-thin components (red and blue). Right panel: Spectrum from day 43.8 after the eruption, around the time the 0.3--1~keV emission peaked, fitted with a blackbody plus edges (magenta), together with two optically-thin components (red and blue).}
\label{spec-model}
\end{center}
\end{figure*}

In both 2006 and 2021, the harder ($>$~1~keV) emission, due to the shocked wind, has been parametrized in this paper by two optically-thin ({\sc apec} in {\sc xspec}) components, fixed at solar abundance. Considering the pre-SSS shock data, there are no strong residuals suggesting non-solar abundances would improve the fits to the {\em Swift} spectra (see left-hand panel of Fig.~\ref{spec-model}).
In some individual observations only one {\sc apec} was required at $\go$~90~per~cent confidence, while others could be improved by the inclusion of a third, cooler component (typically $\sim$~100--200~eV). This is a simplification of what may physically be a more complex underlying shock continuum \citep[see, for example, ][]{vaytet07,vaytet11}. However the two-temperature model accounts for the measured harder emission well. 

Each spectrum was also attenuated by two absorption components: the interstellar medium (ISM) value fixed at 2.4~$\times$~10$^{21}$~cm$^{-2}$ \citep{hjellming86}, and a variable column density due to the RG wind. In \cite{julo11}, this additional wind absorption was constrained to follow a power-law decline over time based on analysis in \cite{bode06}. However, in the current analysis (both for the new 2021 observations, and the re-analysis of the 2006 data), we have chosen to allow this column to vary because of the apparently more complex variation seen with this larger sample of data. The oxygen abundance for the excess N$_{\rm H,wind}$ above the ISM level was fixed at 0.3 solar (within the {\sc xspec/tbfeo} model). This lower oxygen abundance was determined from modelling the grating observations in 2006 \citep{ness07a, julo11}, and typically provides an improvement when fitting the 2021 datasets as well, so was adopted throughout the analysis described below.

The same absorption parameter, N$_{\rm H,wind}$, was applied to all components in each
fit; that is, the soft, optically-thick BB and harder optically-thin components were absorbed by the same column. Although it is possible that the emission from the nuclear
burning on the WD surface and that from shocks with the RG wind
experience different levels of absorption, a single varying column (in
excess of the fixed Galactic value) led to acceptable fits. This excess absorption has been assumed to be neutral.

Fig.~\ref{spec-model} shows two example spectra from the times in 2021 where the 1--10~keV and 0.3--1~keV emission peaked, respectively, with the fitted components shown (red dots and blue dashes for the optically-thin shock emission, and a magenta dot-dash line for the edge-absorbed BB component; in this example, only the He-like oxygen edge was significant). While the models are clearly not perfect, with some fit residuals apparent, the overall fit to the continuum is good (C-stat/dof = 852/766 and 275/245, respectively, for the spectra plotted).

\subsection{Spectral fitting results}

Experimenting with different fits, it became clear that the ionized edges were much more significant in the 2006 spectra, with most of the 2021 soft data being well modelled with a BB alone. Indeed, upon fitting the 2021 spectra with different combinations of edges, it was apparent that there could be significant degeneracy between the edges included and the BB temperature for some of the spectra. That is, fits with a low ($\sim$~50~eV) BB kT and shallow/no edges, or a higher kT ($\sim$~100--150~eV) and deep edges, often lead to a very similar goodness of fit (C-stat values within $\sim$~5 of each other), with little difference in the fit residuals.

\begin{figure*}
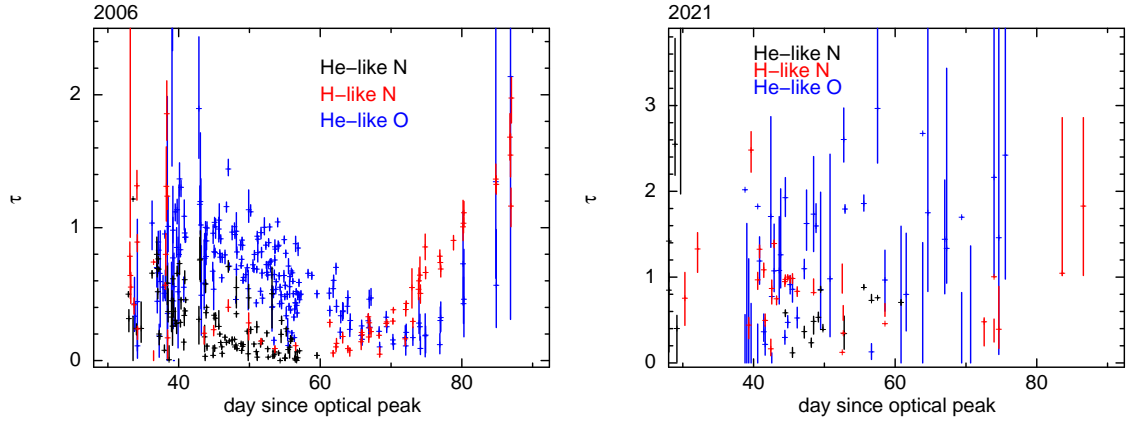

\begin{center}
  \includegraphics[clip, width=5.5cm, angle=-90]{3edgedepths_2006_nozero.ps}
  \includegraphics[clip, width=5.5cm, angle=-90]{3edgedepths_2021_nozero_no5.ps}
  \caption{Plots of the fitted edge optical depths for which the 1$\sigma$ lower limit is inconsistent with zero. For the 2006 dataset (left), the He-like O edge (0.739~keV; blue) was the most significant at early times, decreasing in strength with time. The shallower optical depth of the He-like N (0.552~keV; black) edge followed a similar decreasing trend. H-like N (0.635~keV; red) improved the fits mainly after day 60. The optical depths of the edges fitted to the 2021 spectra (right) do not show any obvious trends. The small number of edges for which the optical depth pegged at the deepest allowed value of $\tau$~=~5 have been omitted from the plots. Note that the vertical scales are different for the two plots.}
\label{tau}
\end{center}
\end{figure*}

The majority of the 2006 spectra were better fitted by including some or all of the four absorption edges described above. The left-hand panel of Fig.~\ref{tau} demonstrates how the optical depths of the edges varied in the 2006 spectral fits; recall that the depth of the H-like O edge at 0.871~keV was fixed at $\tau$~=~2 throughout. Of the three with variable optical depths, the He-like O edge at 0.739~keV was the most significant at early times. This optical depth then decreased, until about day 70, after which time the edge was no longer strongly significant. The 0.552~keV He-like N edge is the second deepest at the start of the SSS phase, also decreasing in strength with time, until becoming insignificant after around day 55. The edge at 0.635~keV, related to H-like N, was really only required after day 60, and its optical depth increased steadily after this time. No such obvious trends were found in the typically less-significant optical depths in the 2021 fits, as can be seen in the right-hand panel of Fig.~\ref{tau}. 
Because of these findings, we choose to show  both single BB fits, and those from a BB plus these four absorption edges (for the SSS emission) as described above; the underlying {\sc apec} components were included to account for the shock emission.

Figs.~\ref{fitsa} and \ref{fitsb} show the results of fitting the spectra with two absorbed optically-thin {\sc apec} components before the onset of the SSS phase, then with a BB without and with absorption edges included as well, for both 2006 and 2021. In both cases, the top panel plots the observed 0.3--10~keV fluxes (in units of erg~cm$^{-2}$~s$^{-1}$) corresponding to the BB component (modelling the SSS; shown in black and red) and the combined {\sc apec} emission (modelling the underlying shocks; grey and blue). The second panel demonstrates the evolution of the temperatures of the two optically-thin {\sc apec} components, while the third does the same for the BB, with the corresponding BB effective radius in the fourth. The bottom panel shows how the excess absorbing column, N$_{\rm H,wind}$, decreases with time. In both plots, the axes in each corresponding panel are scaled to be the same, for ease of direct comparison. The fitted parameters corresponding to these figures are provided in Table~\ref{swiftpar}.

\begin{table*}
  \caption{Fits to the XRT spectra before (model consisting of two absorbed {\sc apec} components) and during (including an additional BB) the SSS phases in 2006 and 2021, as plotted in Figs.~\ref{fitsa} and \ref{fitsb}. The first column lists the mid-point of the spectrum in days since the optical peak of the eruption, while the second column provides the excess absorption. Columns three to five list the temperatures of the BB and {\sc apec} components, while columns six and seven show the observed flux over 0.3--10~keV for the BB and combined {\sc apec} components, respectively. The final column gives the Cash statistic and number of degrees of freedom. The full version of this table is available online.}
  \begin{center}

    \begin{tabular}{lccccccc}
      \hline
  \textbf{T$_{\rm mid}$} & \textbf{N$_{\rm H, wind}$} & \textbf{kT$_{\rm BB}$} & \textbf{{\sc apec} kT$_{\rm cool}$} & \textbf{{\sc apec} kT$_{\rm hot}$} & \textbf{BB Flux} & \textbf{{\sc apec} Flux}& \textbf{Cstat/dof}\\
\textbf{(day)} & \textbf{(10$^{22}$ cm$^{-2}$)} & \textbf{(eV)} & \textbf{(keV)} & \textbf{(keV)} & \textbf{(erg cm$^{-2}$ s$^{-1}$)} & \textbf{(erg cm$^{-2}$ s$^{-1}$)}\\
\hline
\textbf{2006 eruption}\\
3.178	&	6.77$^{+0.40}_{-0.38}$&	---&	0.168$^{+0.023}_{-0.028}$&	8.449$^{+0.637}_{-0.576}$&---	&	1.30$^{+0.02}_{-0.02}$~$ \times$~10$^{-9}$				&	903.6	/	799	\\
5.033	&	2.47$^{+0.10}_{-0.09}$&	---&	0.248$^{+0.017}_{-0.024}$&	7.407$^{+0.312}_{-0.310}$&---	&	2.01$^{+0.02}_{-0.02}$~$ \times$~10$^{-9}$				&	1175.1	/	854	\\
8.183	&	1.68$^{+0.10}_{-0.10}$&	---&	0.280$^{+0.017}_{-0.017}$&	5.964$^{+0.335}_{-0.268}$&---	&	1.11$^{+0.01}_{-0.01}$~$ \times$~10$^{-9}$				&	1007.5	/	781	\\
10.99	&	1.75$^{+0.16}_{-0.20}$&	---&	0.192$^{+0.041}_{-0.028}$&	2.976$^{+0.199}_{-0.165}$&---	&	6.50$^{+0.17}_{-0.18}$~$ \times$~10$^{-10}$				&	655.4	/	602	\\
11.057	&	1.55$^{+0.19}_{-0.20}$&	---&	0.263$^{+0.039}_{-0.054}$&	3.208$^{+0.294}_{-0.251}$&---	&	6.97$^{+0.23}_{-0.21}$~$ \times$~10$^{-10}$				&	563.1	/	574	\\
11.124	&	2.07$^{+0.02}_{-0.08}$&	---&	0.138$^{+0.011}_{-0.007}$&	2.688$^{+0.093}_{-0.082}$&---	&	6.79$^{+0.19}_{-0.12}$~$ \times$~10$^{-10}$				&	574.4	/	585	\\
13.605	&	1.56$^{+0.10}_{-0.11}$&	---&	0.200$^{+0.019}_{-0.012}$&	2.535$^{+0.101}_{-0.074}$&---	&	4.82$^{+0.08}_{-0.08}$~$ \times$~10$^{-10}$				&	749.4	/	659	\\
15.616	&	1.54$^{+0.06}_{-0.06}$&	---&	0.621$^{+0.020}_{-0.058}$&	2.782$^{+0.109}_{-0.096}$&---	&	3.89$^{+0.06}_{-0.06}$~$ \times$~10$^{-10}$				&	890.9	/	639	\\
18.176	&	1.54$^{+0.11}_{-0.13}$&	---&	0.800$^{+0.044}_{-0.038}$&	2.533$^{+0.256}_{-0.203}$&---	&	2.31$^{+0.09}_{-0.04}$~$ \times$~10$^{-10}$				&	408.5	/	437	\\
18.229	&	1.50$^{+0.16}_{-0.18}$&	---&	0.844$^{+0.064}_{-0.070}$&	2.427$^{+0.630}_{-0.380}$&---	&	2.29$^{+0.08}_{-0.10}$~$ \times$~10$^{-10}$				&	257.0	/	311	\\
18.239	&	1.54$^{+0.11}_{-0.13}$&	---&	0.665$^{+0.056}_{-0.051}$&	2.597$^{+0.289}_{-0.212}$&---	&	2.89$^{+0.10}_{-0.10}$~$ \times$~10$^{-10}$				&	447.0	/	461	\\
25.998	&	1.37$^{+0.04}_{-0.05}$&	28.1$^{+1.6}_{-0.8}$&	0.672$^{+0.040}_{-0.015}$&	2.047$^{+0.156}_{-0.127}$&1.11$^{+0.06}_{-0.06}$~$ \times$~10$^{-11}$&	1.70$^{+0.03}_{-0.03}$~$ \times$~10$^{-10}$					&	665.9	/	499	\\
29.01	&	0.94$^{+0.03}_{-0.02}$&	18.1$^{+0.3}_{-0.4}$&	0.701$^{+0.015}_{-0.015}$&	1.569$^{+0.047}_{-0.059}$&2.74$^{+0.02}_{-0.02}$~$ \times$~10$^{-10}$&	1.48$^{+0.02}_{-0.02}$~$ \times$~10$^{-10}$					&	715.8	/	467	\\
29.886	&	0.65$^{+0.03}_{-0.04}$&	25.1$^{+1.0}_{-0.6}$&	0.707$^{+0.038}_{-0.016}$&	1.602$^{+0.056}_{-0.038}$&8.73$^{+0.13}_{-0.13}$~$ \times$~10$^{-11}$&	1.30$^{+0.02}_{-0.02}$~$ \times$~10$^{-10}$					&	836.0	/	432	\\
\hline
\end{tabular}

\label{swiftpar}
\end{center}

\end{table*}

At later times in 2021, there were intervals where the fitted BB temperature was seen to jump dramatically within a day; as discussed above, this is caused by the temperature/edge depth degeneracy. Given that most of the spectra were better fitted  (albeit often only slightly) with the cooler ($\sim$~50~eV) BBs, the sudden apparent jumps to higher temperatures are most probably anomalous, and not physical. For the fits to these spectra, the BB component was therefore fixed at the mean temperature of the two spectra before and after in time where the blackbody temperature was cooler; these data are plotted as magenta stars in Fig.~\ref{fitsb}, with the corresponding cooler/hotter {\sc apec} parameters shown in magenta/cyan. The optical depths of the edges shown in Fig.~\ref{tau} correspond to these lower BB temperature fits.

\subsubsection{Comparison between fits with and without absorption edges}

The main differences seen when absorption edges are included in modelling the observed spectra are an increase in BB temperature (by 10--15~eV), and a decrease in the effective radius of the emitting region (by about a factor of three), in both cases after day 40, once the SSS emission had settled down. 
To a lesser extent, there is also a slight decrease in the excess N$_{\rm H}$ over this same interval. All of these differences are clearer in the 2006 data, where the edges are more strongly detected.

\subsubsection{Comparison between 2021 and 2006 fits}
\label{comp}

The underlying X-ray spectral shape was similar in both 2006 and 2021, with the temperatures and absorbing columns covering approximately the same parameter space, though with certain differences. 

The temperatures of the two optically-thin components, together with N$_{\rm H,wind}$, evolve most strongly during the first 10--15~days after the nova event, both in 2006 and 2021. Beyond this time, the shock emission cools slowly, and the measured absorbing column declines only a little further, although there is a drop in  N$_{\rm H,wind}$ after day $\sim$ 60 in 2021. While a consistently higher absorbing column might present itself as a simple explanation for the lower SSS flux observed in 2021, these fits do not support that hypothesis.

The BB temperatures cover a similar range in both outbursts, starting off cool, around 20--30~eV, before increasing. The inclusion of the edges leads to a larger rise in fitted temperature.
The 2021 data show more variability in the BB parameters at the earliest times (i.e., before about day 30), when the SSS is only weakly present; this is particularly the case where absorption edges are included in the fit, which is likely a symptom of over-fitting the data. 


\begin{figure*}
\begin{center}
\includegraphics[clip, width=15cm]{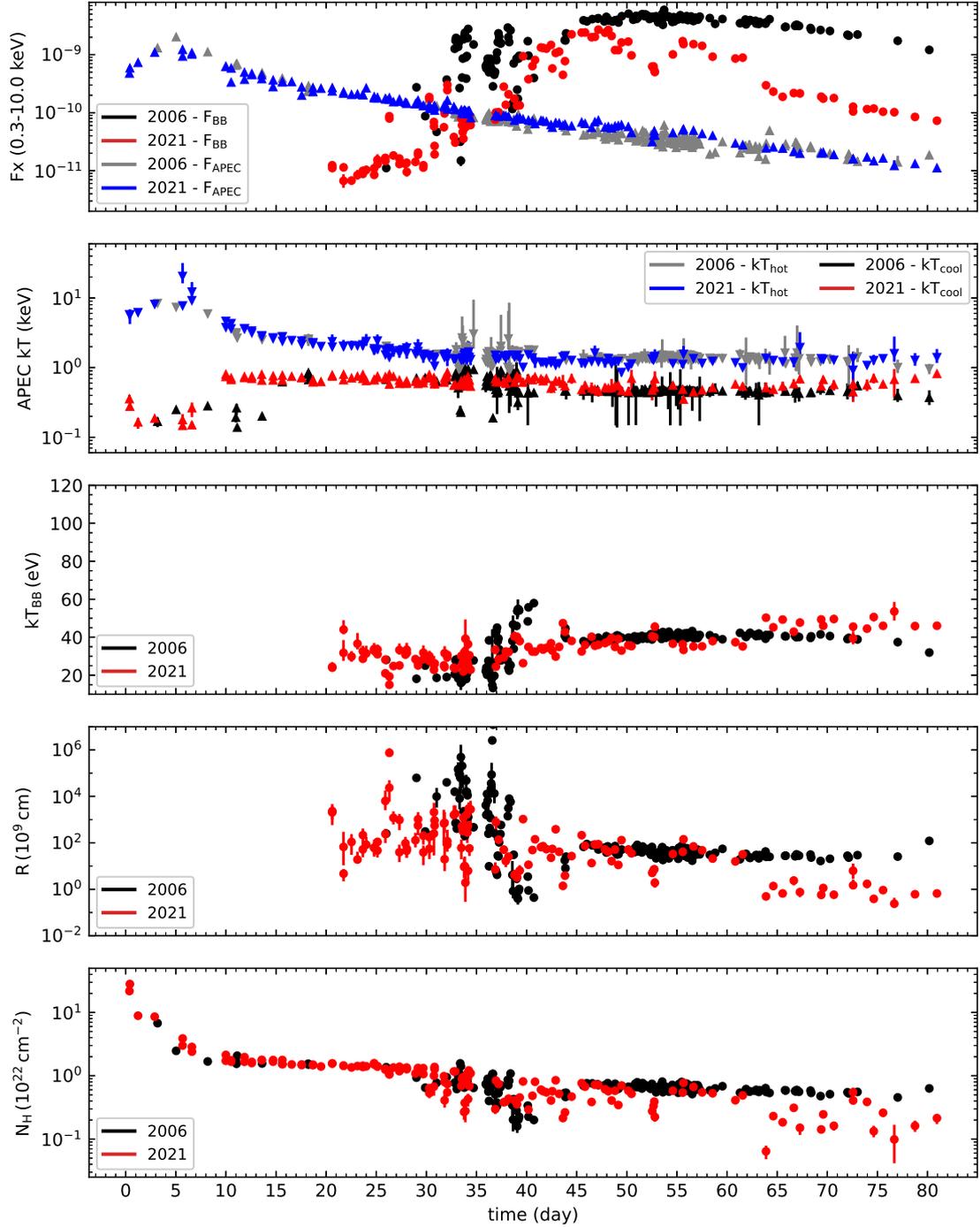}  
\caption{Results from fitting the spectra from 2006 and 2021 with a simple BB to model the soft emission, together  with two optically-thin {\sc apec} components. The top panel shows the observed 0.3--10~keV fluxes from the BB and {\sc apec} components (in units of erg~cm$^{-2}$~s$^{-1}$). The second panel shows the evolution of the {\sc apec} component temperatures, while the third panel shows the same for the BB. The fourth panel shows the emission radius for the BB component, and the bottom panel shows how the absorbing column N$_{\rm H,wind}$ (in excess of the Galactic ISM value) evolves with time. 2006 data are plotted in monochrome, while 2021 are in blue (hotter {\sc apec} parameters) and red (cooler {\sc apec} and all other parameters). Note that the ordinate scales are the same in each panel as in Fig.~\ref{fitsb}, for ease of comparison. See text for more details of the models.}
\label{fitsa}
\end{center}
\end{figure*}

\begin{figure*}
\begin{center}
  \includegraphics[clip, width=15cm]{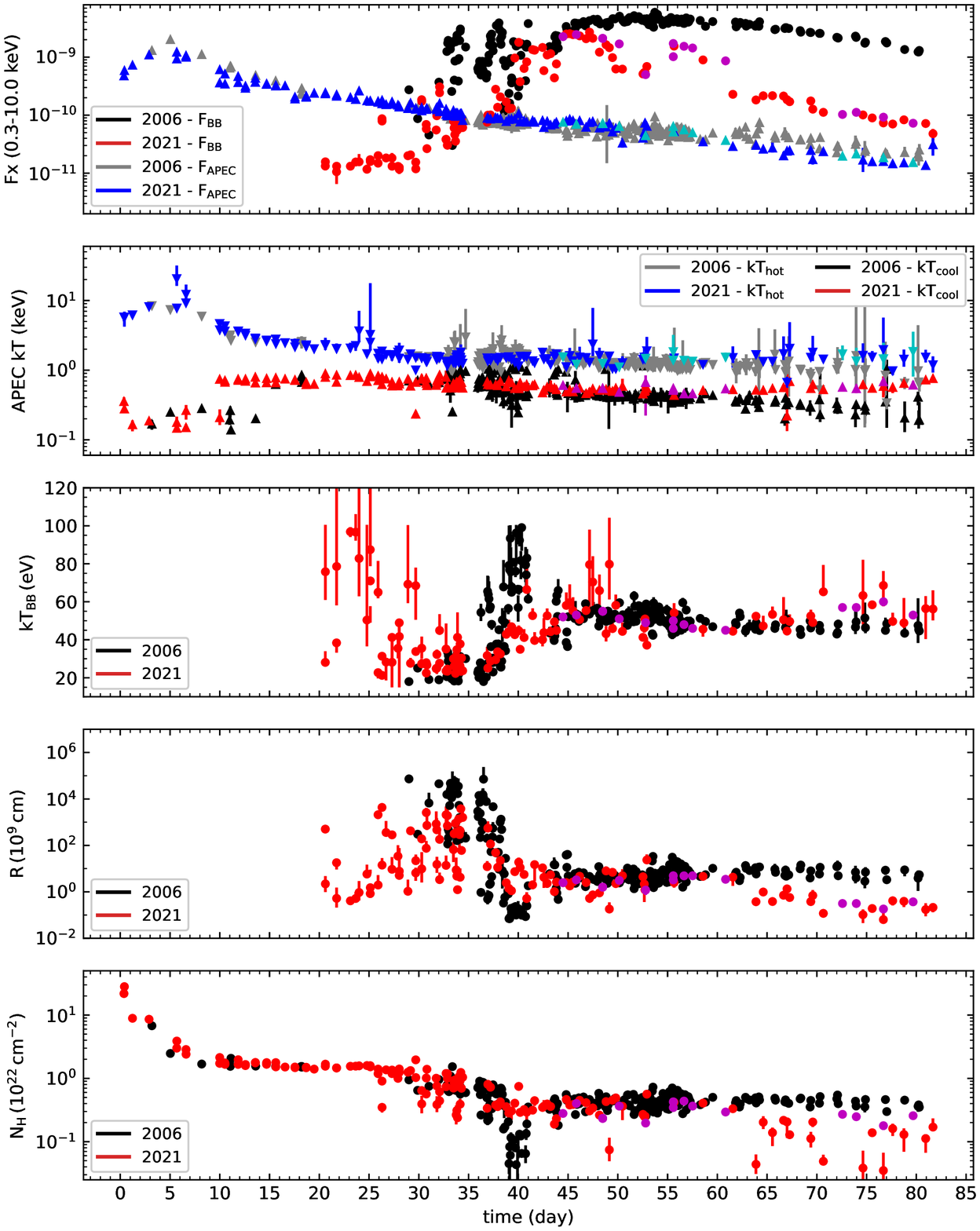}
  \caption{As for Fig.~\ref{fitsa}, but including four absorption edges as well.
2006 data are plotted in monochrome, while 2021 are in blue/cyan (hotter {\sc apec} parameters) and red/magenta (cooler {\sc apec} and all other parameters: red where BB kT was free to vary, magenta stars where that temperature was constrained. The {\sc apec} temperatures plotted in magenta and cyan correspond to the magenta BB bins, likewise for the radius and N$_{\rm H}$.) Note that the ordinate scales are the same in each panel as in Fig.~\ref{fitsa}, for ease of comparison. See text for more details of the models.}
\label{fitsb}
\end{center}
\end{figure*}

A large difference between the datasets is the excursion in BB kT around day 40 in 2006 (with and without absorption edges), where these data are better fitted with a temperature about 20--40~eV higher than observations shortly before or after this time (a bigger discrepancy when the edges are included). N$_{\rm H,wind}$ is correspondingly lower during this interval.  The fitted temperature and column density are, of course, correlated to some level, due to the modest spectral resolution of the XRT at these energies. Forcing a higher N$_{\rm H,wind}$ causes the BB temperature to decrease (or vice versa); these forced fits are statistically worse, though. These observations with higher BB kT occurred during the time of large-scale flux variability -- something which was not seen to such an extent in 2021, so it is perhaps unsurprising that the same variation in temperature/absorption is not found following the latest eruption.

\cite{julo11} describe a rather different evolution of the SSS temperature compared with the re-analysis of those same 2006 data herein, finding significant changes in kT throughout the early flux variability phase, which then settled at the highest value reached. This may be caused by the differences in how the absorption is modelled. As noted above, \cite{julo11} constrain N$_{\rm H,wind}$ to follow a power-law of t$^{-0.5}$~cm$^{-2}$ after the first detection of the SSS emission \citep[based on work in][]{bode06}. In our re-analysis, where N$_{\rm H,wind}$ is freely fitted, we initially find a more variable column, particularly during the chaotic start to the SSS phase (though note mention of N$_{\rm H,wind}$-BB kT degeneracies above). However, after $\sim$~day 50, the fit results for the 2006 dataset are not inconsistent with a power-law decline in N$_{\rm H,wind}$ of index 0.5, although there is still significant scatter. The estimate of the range of the absorbing column in 2021 is very similar to that in 2006 until around day 60, when it decreases more significantly.

Around day 53 in 2021, there seems to be a brief dip in the effective BB radius, and an upward step in the temperature, corresponding to the temporary drop in observed soft flux at that time. About a week later, after day 60, the 2021 BB temperatures become consistently slightly hotter, with a notable decrease in the corresponding emitting radius compared with 2006. It is not unexpected that, as the WD runs out of fuel, the bloated outer layers begin to shrink and, consequently, the BB temperature rises \citep[e.g., ][]{macdonald85, krautter96, shore96b}. However, from our fit results, the drop in radius is only obvious in 2021 over the time frame considered (i.e., before day 86). This could be an indication that the nuclear burning phase ended earlier in 2021 -- though see discussion in Section~\ref{fe}. The geometry of nova systems post-eruption is not always clear, though, and the WD may remain bloated long after the end of the eruption \citep[e.g.,][]{mason21}.

\begin{figure}
\begin{center}
  \includegraphics[clip, angle=-90, width=8cm]{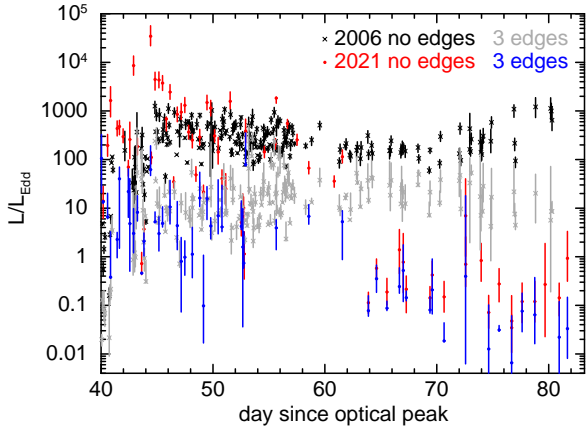}
  \caption{Bolometric luminosity estimates from the BB fits for the 2006 (black/grey) and 2021 (red/blue) data, without and with absorption edges. Whether or not the edges are included, the 2021 data are systematically less luminous than the 2006 data after day 60. The luminosity is plotted in terms of the Eddington value for a 1.4~$\msun$ WD (1.76~$\times$~10$^{38}$~erg~s$^{-1}$), commensurate with the upper limit on the WD mass in \citet{miko17}.}
\label{lum}
\end{center}
\end{figure}

In Fig.~\ref{lum} we show the bolometric luminosities estimated from the BB fits to the data after day 40, for both the zero and multiple edge models, in 2006 and 2021. The inclusion of (significant) absorption edges decreases the estimated luminosities overall, in both years. The low 0.3--10~keV flux measurements seen in 2021, particularly after day 60, clearly correspond to much lower bolometric luminosities. This is also the time frame over which the BB radius decreases (consistently) in 2021. It is well-known that BB models can overestimate luminosities \citep[e.g.,][]{krautter96}, so the high peak values of 100-1000~L$_{\rm Edd}$ shown in Fig.~\ref{lum} should not be taken as physically realistic, and only the comparative trends considered. 

The optical [Fe\,{\sc x}] 6375~\AA\ coronal line (ionization energy of 235~eV) is often considered to be a tracer for the SSS phase, since it acts as a good indicator of photoionization from a hot WD \citep{krautter89}. While more energetic photons from the shocked ejecta will also be able to excite the [Fe\,{\sc x}] emission line, the behaviour of the line is much more reminiscent of the SSS evolution.

\begin{figure}
\begin{center}
  \includegraphics[clip, width=8cm]{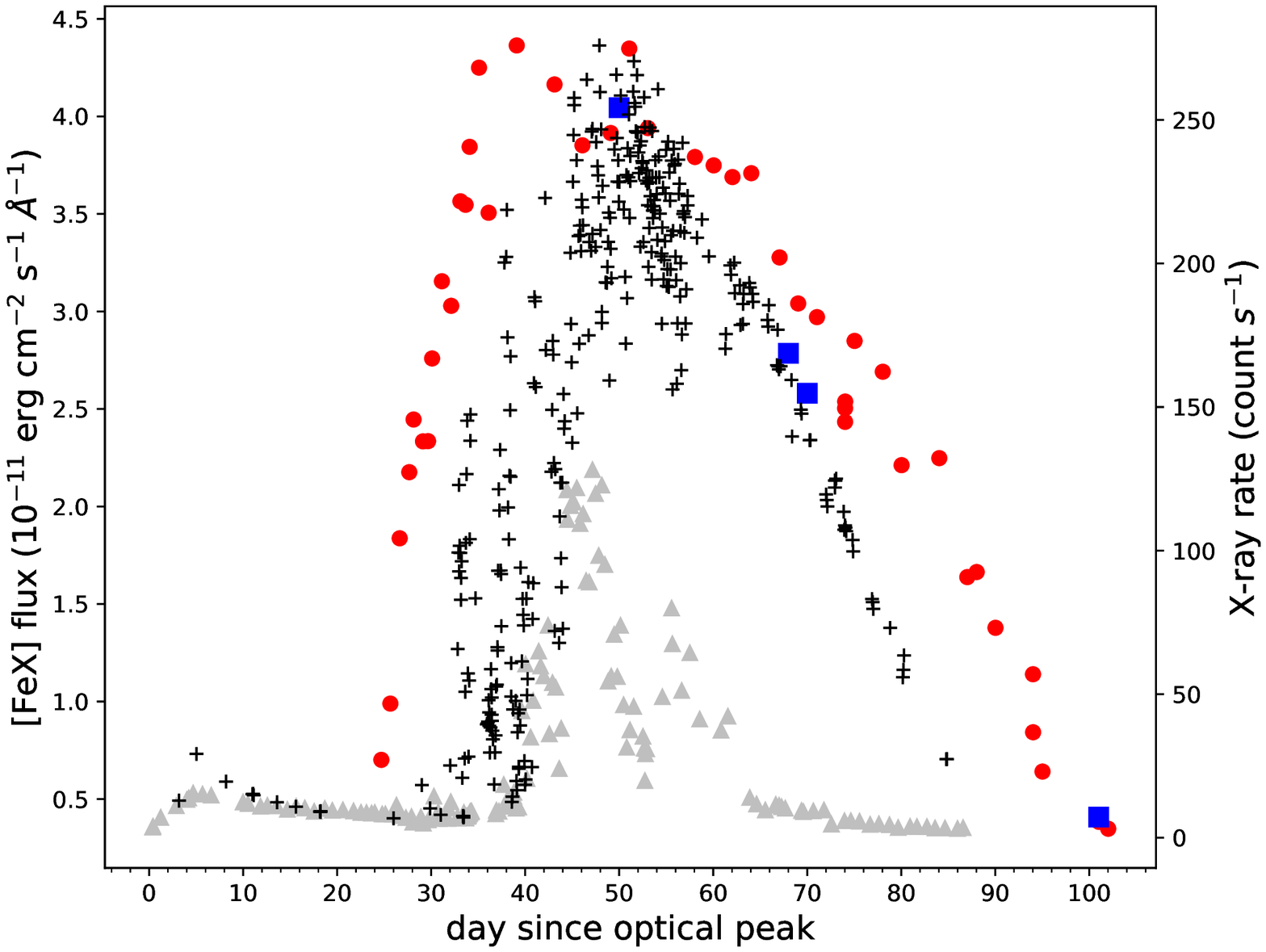}
  \caption{The evolution of the [Fe\,{\sc x}] flux in 2021 and 2006 (red circles and blue squares; left-hand axis), compared with the corresponding 0.3--10~keV X-ray light-curves (grey triangles and black pluses; right-hand scale).}
\label{fex}
\end{center}
\end{figure}

The evolution of the optical [Fe\,{\sc x}] line flux is therefore presented in Fig.~\ref{fex}, with the X-ray light-curves for comparison.
In 2021 the last non-detection of the line occurred on day 23.10, with the first positive detection occurring on day 25.10. In an effort to compare with the previous 2006 outburst, we have also measured the flux of [Fe\,{\sc x}] on four of the eight spectra presented by \cite{munari07} -- those for which an accurate absolute flux calibration is available -- and these results for 2006 are over-plotted as blue squares in Fig.~\ref{fex}. The overall shape and duration of the [Fe\,{\sc x}] curve is more similar to the X-ray evolution seen in 2006 than in 2021.

\section{1985 eruption}

Following the 1985 eruption, {\em EXOSAT}\footnote{European X-ray Observatory Satellite} observed RS~Oph six times between days 54 and 250 after the peak of the optical emission, the first time the nova had been observed in X-rays \citep{mason87}. Four of these observations lie within the post-outburst time frame we are considering in this paper. We extracted the count rates from the HEASARC database\footnote{\url{https://heasarc.gsfc.nasa.gov/cgi-bin/W3Browse/w3browse.pl}}, and converted the LE-LX3 (Low Energy imaging telescope with the Lexan 3000 filter\footnote{The Medium Energy proportional counter (ME; 1.2--50~keV) was also used, but the source was only detected during the first three observations.}) measurements (0.05-2~keV) to {\em Swift} 0.3--1~keV count rates using PIMMS\footnote{Portable, Interactive Multi-Mission Simulator: \url{https://heasarc.gsfc.nasa.gov/cgi-bin/Tools/w3pimms/w3pimms.pl}}. This gave an approximate conversion of 1~count~s$^{-1}$ with the LE-LX3 corresponding to $\sim$~13--17~count~s$^{-1}$ in XRT-WT (grade 0) over 0.3--1~keV; this range was estimated by taking the BB and N$_{\rm H}$ values from the fits closest in time to each {\em EXOSAT} post-eruption day from both 2006 and 2021\footnote{We note that \cite{ness07b} scaled the {\em EXOSAT} data by eye to reproduce the XRT count rate on day 55 after the 2006 eruption, leading to a very different comparison.}. \cite{mason87} attempted to fit the {\em EXOSAT} ME data over $\sim$~1.5--5.5~keV. They found that the spectrum shows a strong rise at the soft energy end, indicating the emission was likely super-soft, as assumed for this conversion.

\begin{figure}
\begin{center}
  \includegraphics[clip, angle=-90, width=8cm]{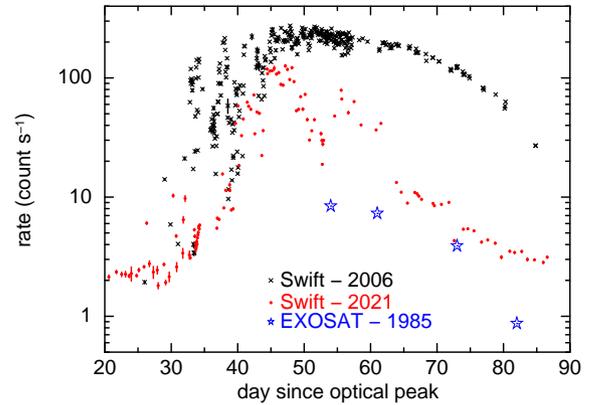}
  \caption{Comparison of the {\em EXOSAT} LE/LX3 observations following the 1985 eruption of RS Oph with the {\em Swift} light-curves from 2006 and 2021. The {\em EXOSAT} measurements have been converted to {\em Swift}-equivalent 0.3--1~keV count rates using WebPIMMS.}
\label{exosat}
\end{center}
\end{figure}

Fig.~\ref{exosat} shows these converted {\em EXOSAT} measurements (using the mean conversion of 1 {\em EXOSAT}-LE count~s$^{-1}$ = 15 {\em Swift}-XRT count~s$^{-1}$) with respect to the {\em Swift} light-curves from 2006 and 2021, indicating that the soft X-ray emission in 1985 was even fainter than in 2021. However, the 1985 data drop off rapidly after day 60--70, more like the 2006 eruption than the 2021 one. Interestingly, the 1985 and 2006 eruptions occurred at about the same binary phase, while 2021 was $\sim$~180$^o$ different. Note that the RS~Oph eruption prior to 1985 occurred in 1967, so the 1985 eruption followed a quiescent interval of 18~yr.

The X-rays detected with the limited {\em EXOSAT} observations were previously interpreted to be shock emission \citep{mason87}, although \cite{obrien92} suggested that a better solution might be that the shock in the RG wind provides the higher-energy X-ray emission whilst most of the low-energy flux is due to the white dwarf remnant. 
With the more detailed observations from {\em Swift}, we can state that the majority of these X-rays were most likely supersoft photons from the surface nuclear burning as it switched off \citep[as also concluded by][]{ness07b}. Indeed, while \cite{mason87} interpreted the earlier observations (including all those presented here in Fig.~\ref{exosat}) as shock-heated thermal emission, they ascribe the last detection around day 250 to residual nuclear burning on the WD surface.

\section{Discussion}

With a mean recurrence timescale of 15~yr, RS~Oph is the first Galactic recurrent nova for which more than one eruption has been monitored by {\em Swift}. The quiescent interval preceding the 2021 eruption was 15.5~yr, close to this mean value; the 21~yr gap from 1985--2006, however, is around 35~per~cent longer than the average. In this paper we perform a detailed comparison of the XRT data collected during the first $\sim$~86~days of these eruptions, to investigate how similar -- or different -- the events were.

The most obvious difference which requires explanation is the clearly brighter observed X-ray emission in 2006 compared with 2021 (and 1985). Here we summarise all the salient points:

\vspace{5mm}
{\bf Similarities}:
\begin{itemize}
\item The optical light-curves in 2006 and 2021 are almost identical. This strongly suggests that the WD mass and ejecta mass, velocity distribution and geometry are close to constant, since these are the parameters which affect the optical evolution \citep[e.g., ][and references therein]{shore12, chomiuk21}.
\item Despite limited data from 2006, the evolution of the [Fe\,{\sc x}] coronal emission appears to be approximately the same following both eruptions.  
\item  The harder X-ray emission (dominating $>$~1~keV) produced by shock interactions between the nova ejecta and the RG wind was similar, in terms of both brightness and temperature evolution, throughout the majority of the interval considered. This, like the identical optical light-curves, implies the ejecta parameters are similar, as well as the gross properties of the RG wind density structure close in to the system.
\item The SSS emission became visible around the same time (day 20--26), and the final decline started around day 60 after each eruption.
\item The magnitude and temporal evolution of the wind absorbing column (in excess of the Galactic value) was typically similar until day 60, although the 2006 data show more variation during the end of the high-amplitude flux changes (around day 40).
\item The parametrization of the SSS with a BB mainly provided similar temperatures between the two eruptions, with or without the inclusion of absorption edges. 
 \item A QPO was identified in both 2006 and 2021, with a consistent mean period of $\sim$~35~s. The modulation fraction was also similar during both eruptions. 
\end{itemize}

{\bf Differences}:
\begin{itemize}
\item The measured count rate at the beginning of the SSS phase in 2006 was more variable than in 2021.
\item The count rate peaks and troughs occurred at different times following each eruption.  
\item The SSS emission in 2006 was brighter than in 2021, peaking at more than double the count rate over 0.3--1~keV. The integrated number of soft X-ray counts is $\sim$~4--5 times larger in 2006 than 2021.
\item The 2006 eruption occurred after an inter-eruption gap of 21~yr, about 35~per~cent longer than the average recurrence time. The 15.5~yr quiescent interval which ended in 2021 is almost exactly the average duration.
\item In 2006, the SSS emission plateaued around a peak count rate between $\sim$~days 45--60, before a monotonic decline set in. The 2021 soft emission stayed at peak brightness from around days 45--48, then faded by a factor of about five over the next five days, followed by a brief rebrightening before a final fading trend set in around day 60.
\item After day 76, the rate of decline of the 2006 soft emission was more than twice as rapid as at the corresponding time in 2021.
\item As noted above, the range of BB temperatures was similar during both SSS phases. However, the 2006 spectra show a more rapid increase in temperature, then staying approximately constant until day 70. The 2021 temperature increase is gradual, becoming systematically hotter than the corresponding 2006 data after day 60; this is more obvious with the simple BB fits with no edges included, but the trend is still there with the edges. The 2006 data are also noticeably hotter for an interval around day 40.
\item The 2006 spectra show more signs of absorption features, with the 2021 spectra appearing much smoother; those fits are therefore not significantly improved by the inclusion of ionized absorption edges, whereas the 2006 data are.
\item After $\sim$~day~60, the radius of the emitting region in 2006 is consistently higher than in 2021, caused by the 2021 radius dropping, while the 2006 measurement stays approximately constant. This is the case with or without absorption edges. There is also a brief decrease in the 2021 BB radius around day 53, corresponding to the temporary drop in soft flux, though this is not clearly seen when edges are included.
\item N$_{\rm H,wind}$ is also lower after $\sim$~day~60 in 2021, though very similar in both 2006 and 2021 at earlier times.  
\item After day 60, the apparent bolometric SSS luminosity in 2006 was much higher than in 2021.

\end{itemize}

The results of the spectral fitting (Figs.~\ref{fitsa} and \ref{fitsb}) imply the higher SSS count rate and luminosity in 2006, particularly after day 60, were due to a larger effective emitting surface compared with the recent 2021 eruption. There is no indication that the lower count rates in 2021 are caused by higher absorption, though we return to this possibility in Section~\ref{abs}. From the bullet points above, it is notable that many of the differences between the 2021 and 2006 X-ray spectral evolution results occur after day 60 -- that is, the time both X-ray light-curves showed a final monotonic fading, signifying the end of the nuclear burning. At earlier times, during the onset of the SSS phase, there is also a significant difference in the level of soft X-ray variability seen.

\subsection{Differences in SSS related to quiescent interval?}

While the {\em EXOSAT} dataset is much smaller than those of {\em Swift},
it is clear that there were very significant differences in the soft X-ray emission between the three RS~Oph eruptions, which followed quiescent intervals of around 18, 21 and 15.5~yr, respectively. M31N~2008-12a, the yearly recurrent nova in the Andromeda Galaxy, has been observed by {\em Swift} during every eruption since 2012. For the first four years of observations, the X-ray light-curves of M31N~2008-12a were very similar; the 2016 event was a surprise, though, showing a delayed start leading to a shorter and less luminous SSS phase \citep{henze18}. 
The optical light-curve, however, showed little variation from outburst to outburst, which is also the behaviour witnessed for the RS~Oph optical data. The inter-eruption time for M31N 2008-12a varies by $\sim$~15~per~cent, year on year \citep{darn16}, while still (except for 2016) showing a consistent SSS phase.
\cite{henze18} suggest that the differences seen in 2016 (delayed eruption -- i.e., longer quiescent interval; earlier decline in the SSS emission) could be explained by a lower accretion rate, $\dot{M}_{\rm acc}$, between outbursts. This would have led to a lower mass disc, which was thus more disrupted by the nova explosion than the discs usually formed. A more thoroughly disrupted disc in turn leads to a longer delay before effective accretion restarts, hence shortening the SSS phase in this system. In the current case of RS~Oph, the 2006--2021 quiescent interval was {\it shorter} than 1985--2006; however, the 2021 SSS was nonetheless fainter. Thus the behaviour of RS~Oph does not replicate the longer quiescent interval/fainter SSS situation seen in  M31N~2008-12a. There is also no strong sign of the mean accretion rate being different during the RS~Oph inter-eruption intervals (see Section~\ref{quiescence} below).

\subsection{Quiescence}

\subsubsection{Accretion rate and luminosity}
\label{quiescence}

Modelling by \cite{prialnik82} suggests that different accretion rates, $\dot{M}_{\rm acc}$, can lead to different accreted masses being required to initiate a TNR. If $\dot{M}_{\rm acc}$ is higher, both the accretion energy, and the rate at which the compressional energy is released, increases. This causes the temperature on the WD surface to increase more rapidly for a given amount of accreted mass, meaning that less mass needs to be accreted before the environment becomes hot enough to initiate a TNR. In such a situation (i.e., higher $\dot{M}_{\rm acc}$ over a shorter interval), we would expect a weaker nova explosion, with less material available to burn. Thus, if the accretion rate were higher during the 2006--2021 interval compared with 1985--2006, we could expect a less extreme eruption, with a shorter/weaker SSS phase in 2021.

RS~Oph was intermittently observed by {\em Swift} between the 2006 and 2021 eruptions. Considering the 31.3~ks of data collected between 2009 Aug. and 2020 Jul. (days 1284--5269 after the 2006 eruption, during which time the X-ray count rate remained close to constant), the spectrum can be approximated with an optically-thin model, with kT~=~1.0~$\pm$~0.2~keV (C-stat/dof = 55/69). This spectrum has a 0.3--10~keV count rate of 4.0~$\times$~10$^{-3}$~count~s$^{-1}$ (well below the level observed just before the Sun constraint in 2021), which corresponds to a bolometric (unabsorbed) luminosity of $\sim$~6~$\times$~10$^{31}$~erg~s$^{-1}$. Such a measurement is in the same range as the quiescent luminosity determined from {\em ROSAT} observations between the 1985 and 2006 outbursts (\citealt{orio93}; see also estimates in \citealt{mukai08}). This suggests a consistent mass accretion rate during both quiescent periods -- though is by no means definitive, given the small number of observations being compared.

A comparison of the AAVSO optical light-curves from 1985--2006 and 2006-2021 also shows no large differences during the inter-eruption periods. The optical magnitude does appear to brighten by up to a magnitude during the $\sim$~5~yr before each eruption, though. The mean quiescent visual magnitudes from these datasets are 11.3 (1985--2006) and 11.1 (2006-2021), each with a standard deviation of $\sim$~0.4. However, in general the disc luminosity is proportional to $\dot{M}_{\rm acc}$. A $\sim$~35~per~cent longer recurrence time up to 2006 would then suggest $\dot{M}_{\rm acc}$, and hence the disc luminosity, are $\sim$~35~per~cent lower, which equates to a change in magnitude of about 0.3. This would likely be swamped by the optical light from the donor star, so measuring such a change in the accretion rate in this manner would be tricky. Observations in the near UV band should be more informative.

\cite{julo11} comment that this estimated quiescent X-ray luminosity is orders of magnitude lower than the expected inter-outburst accretion value of $\sim$~few~$\times$~10$^{36}$~erg~s$^{-1}$.
\cite{mukai08} consider methods through which the expected quiescent X-ray emission could be suppressed. They concluded it unlikely that absorption could be the sole cause, since an unfeasibly high column would be required to hide emission over 2--10~keV. If the boundary layer between the accretion disc and WD surface were optically-thick, then softer X-rays would be emitted which could more easily be strongly absorbed, although an optically thick boundary layer is not expected in quiescence. \cite{nelson11} analyse observations obtained by {\em Chandra} and {\em XMM-Newton} 537--744 days after the 2006 eruption, finding the X-ray spectra can be modelled with a two-component plasma model (as we see here, underlying the SSS). Using this model, they determine an intrinsic accretion luminosity of 1--2~$\times$~10$^{35}$~erg~s$^{-1}$ from the boundary layer. They find that a combination of a complex (partial-covering) absorber and optically-thick emission from the boundary layer could indeed account for the apparently low quiescent luminosity measured. The X-ray flux at this time was still apparently following a power-law decline which started around day 100 post the 2006 eruption \citep{julo11,page20}, although considering the later light-curve as well, there were signs that the decline might have been ending; thus the interpretation of the flux as accretion-powered may not be secure.  

To summarise: the measured X-ray emission during quiescence provides no strong evidence for a difference in accretion rate between 1985--2006 and 2006--2021. A simple estimate of the accretion luminosities over these two quiescent intervals gives values orders of magnitude lower than expected, though this conflict might possibly be resolved if there is complex absorption in the system.

\subsubsection{Accreted mass}

The accreted mass required to trigger the TNR in RS~Oph is
$\approx$~4.4~$\times$~10$^{-6}~\msun$ (\citealt{julo11}, following \citealt{truran86}), and a steady mass transfer over an interval of 21 yr will obviously lead to a greater build up of
material than over only 15.5~yr. Although having less accreted material to burn might account for a fainter, or shorter-lived, SSS phase in 2021, it fails to explain why the nova did not erupt 5.5~yr before the 2006 event, when that same amount of material had been built up. Neither does it explain the fact that the optical evolution is identical, or that the turn-on time of the SSS phase is very similar.
As discussed above, \cite{prialnik82} show that a higher accretion rates might lead to lower accreted masses being needed to trigger a TNR, but there is no strong evidence that $\dot{M}{_{\rm acc}}$ has varied significantly between the recent eruptions of RS~Oph.

\cite{julo11} find that the mass burnt/ejected during the 2006 nova is likely only a few percent of the accreted envelope, and that the WD in the RS~Oph system is therefore gaining mass with time \citep[see also][]{hachisu07}. This mass gain during a single eruption cycle is, however, relatively small, and will not noticeably change the expected recurrence time.

\subsection{[Fe\,{\sc x}]}
\label{fe}

As noted above, the [Fe\,{\sc x}] line  is thought to mirror the evolution of the central source luminosity, and is therefore suggested to match up with the SSS emission seen in the X-ray band \citep[see discussion in][]{schwarz11}. From Section~\ref{soft}, the first hint of a soft excess in the 2021 X-ray spectrum was day 20.6, becoming significantly stronger from day 26.3. A monotonic decline in the X-ray flux set in after about day 60. These timings are loosely replicated by the changes in the [Fe\,{\sc x}] flux.

While there are many fewer measurements available from 2006 than 2021, the [Fe\,{\sc x}] fluxes in Fig.~\ref{fex} align fairly well following the two eruptions. The 2021 [Fe\,{\sc x}] curve is much smoother than the soft X-ray emission we measured in 2021 (see also Fig.~\ref{lc}), and more reminiscent of the underlying shape of the 2006 X-ray light-curve. The [Fe\,{\sc x}] emission is believed to come from the body of the ejecta photoionized by the WD, and should generally be insensitive to occulting blobs of matter which would absorb the soft X-rays. The [Fe\,{\sc x}] evolution shown in Fig.~\ref{fex} suggests that the nuclear burning was again relatively steady throughout the 2021 eruption, and likely very similar in duration in both 2006 and 2021. This therefore does not seem to agree with the inferred decrease in the BB radius in 2021 corresponding to an earlier end to the nuclear burning, but rather suggests the SSS phase was the same both years.

Note that the recombination timescale of [Fe\,{\sc x}] in this density environment will likely be a day or less \citep[][Munari et al. in prep]{opt5}, so cannot cause the differences seen.

Fig.~5 of \cite{munari22} shows that the fluxes of the coronal lines [Fe\,{\sc x}], [Fe\,{\sc xi}] and [Fe\,{\sc xiv}] all begin to decline from their plateau phase around day 87 (seen as a steepening in the decay in Fig.~\ref{fex}, where we show the [Fe\,{\sc x}] flux in linear space), which strongly suggests the same power source (presumably the nuclear burning) switched off for them simultaneously, at which point the WD would start to cool. The shock breakout at the edge of the RG wind could also have an effect, however \citep{shore96a}.

\subsection{Absorption}
\label{abs}

Looking solely at Fig.~\ref{lc}, one's first assumption might be that the lower soft flux in 2021 is due to greater absorption than in 2006.  Excluding the interval of high-amplitude flux variability between days 30 and 40 \citep[which may itself be caused by clumpy ejecta;][]{julo11}, the 2021 observed flux is systematically much lower than in 2006 after day 45. This drop would require a large absorbing structure moving into view at this time, given that the 2021 flux is significantly lower for at least 40 days (about 10~per~cent of the binary orbit). If absorption were the cause of the 1985 X-rays being fainter, then the absorbing structure would need to be non-permanent, since the 1985 and 2006 eruptions occurred at similar orbital phases.

The simulations by \cite{booth16} show a spiral accretion wake built up during the quiescent mass transfer, potentially positioned along our line of sight to the WD during the time of the 2021 eruption (around phase 0.72). In comparison, the 2006 nova took place at phase 0.26, where the simulations show a lower density. The 1985 eruption occurred around phase 0.32, closer to the 2006 viewing angle than 2021, but the X-ray emission measured in 1985 was even fainter than in 2021. \cite{booth16} do find their simulated structure to be clumpy, though, which could possibly account for this. Modelling by \cite{orlando09, drake09, walder08} also shows distinctly aspherical mass loss in the binary system, all indicating the absorption seen will likely depend on the phase of the orbit during the observations.

\cite{shore93} and \cite{shore96a} investigated the UV spectra of recurrent novae, including RS~Oph following the 1985 outburst, finding differential absorption along the line of sight, and that the orbital modulation of the intensity of the ionizing source by the circumstellar medium can affect the strength of the emission lines. That is, no change in the photoionization source itself was needed to cause the spectral changes seen; rather, an aspect change in the line-of-sight opacity could explain the differences. 

Considering the more recent eruptions of RS~Oph, work by \cite{azz21} and Azzollini et al. (in prep.) finds that, although the optical photometric light-curves from 2006 and 2021 are the same, individual line fluxes (measured from the UVOT UV grism spectra) in 2021 are a factor of $\sim$~10--100 below the corresponding values in 2006. There is a chromatic dependence to this, with shorter wavelength lines showing larger decrements; this may be due to Rayleigh scattering -- the interaction of photons with bound electrons. That is, the neutral hydrogen, distributed around the emitting source, scatters the optical and UV line photons out of our view. Thus, if, in 2006, our line of sight passed through a more ionized region (for example, if the ionization of the wind is variable at different orbital phases), less Rayleigh scattering would occur, and higher line fluxes would be measured -- which is indeed the case.

Whatever the simulations and modelling potentially show, however, the results from our spectral fitting do not obviously support significant differences in absorption. Figs.~\ref{fitsa} and \ref{fitsb} show that the fitted N$_{\rm H,wind}$ is largely consistent between the two eruptions, with the fainter 2021 data actually showing a {\em lower} column after day 60. This drop in N$_{\rm H,wind}$ after day 60 in 2021 could conceivably be related to the edge of the RG wind, established since the 2006 eruption. With the shorter recurrence interval, the wind has only had 15.5~yr of travel time, rather than the 21~yr prior to 2006. In 2006 this wind shock breakout time was found to occur around day 80 \citep{anupama08}.

We considered a small sample of spectra where we have data on the same day post-eruption in both 2006 and 2021. In each case, the best-fit 2006 model was applied to the corresponding 2021 spectrum, with only the absorbing column density and edge depths allowed to vary. This assumes a simplified model, whereby the underlying SSS and shock emission were exactly the same at corresponding days in both 2006 and 2021, and the only difference was the intervening column. The resulting fits were statistically much worse ($\Delta$~C-stat $\sim$ 100--700), indicating these models are not acceptable, and it is unlikely we are seeing identical nuclear burning (temperature and strength), with only a change in N$_{\rm H,wind}$. The fit results in Figs.~\ref{fitsa} and \ref{fitsb}  also imply that the emitting radius was smaller after around day 60 in 2021, though it is not clear whether this is caused by the ending of nuclear burning, or some other effect. 

The absorption fitted to the X-ray spectra presented here has been assumed to be neutral. If, instead, there is an ionized component as suggested by the optical/UV analysis, or the abundances are non-solar, this could affect the results. Indeed, the ionized He-like and H-like nitrogen and oxygen absorption edges are notably more significant in the 2006 X-ray spectral fits, which could be a sign of the differing ionization states observed.

In summary, we find that the lower soft X-ray flux seen in 2021 compared with 2006 cannot readily be explained by an increase in simple neutral absorption. The investigation by \cite{nelson11} of post-eruption RS~Oph data suggests that the absorbing column(s) may be more complex, however. We reiterate that the inclination angle of the RS~Oph system is estimated to be in the range 30--52$^{o}$, which places constraints on the amount of absorption to which the accretion disc could lead.

A paper analysing the much higher signal-to-noise {\em NICER} spectra is in preparation by Orio et al., and this may shed more light on the spectral complexities.

\subsection{A Combination Nova?}
\label{comb}

The outbursts of RS~Oph have been suggested to be due to a dwarf nova (DN) disc instability, in which a burst of enhanced mass accretion triggers the outburst, rather than a TNR due to steady accretion during quiescence \citep{king09}, although a TNR-driven outburst is preferred \cite[see summary by ][for example]{obrien92}. 

A related suggestion is that RS~Oph is a `combination nova' \citep{sok06}. These are sources which show both dwarf and classical nova characteristics: eruptions in these systems start off as an accretion disc instability (as happens in DNe), leading to sufficient material being accreted to trigger the thermonuclear burning and mass ejection expected from a `normal' nova. \cite{alexander11} suggest that the outbursts in RS~Oph could be powered by an accretion dump due to a single disc instability, while the model by \cite{bollimpalli18} requires multiple DN events to provide the fuel more slowly to cause a TNR over the recurrence time observed. These disc instability events should be detectable in X-rays, yet no such events have been reported. However, dwarf nova outbursts can be faint, and RS~Oph is more distant than most DNe \citep[e.g., ][]{patterson11}. We also note that the model by \cite{bollimpalli18} requires the WD to have a significant magnetic field, as we discuss for the QPO.

Could it therefore be that, in 2006, a disc instability led to a greater amount of material being dumped onto the WD than was the case in 2021 (or 1985)? This might explain the overall brighter SSS emission, and the fact that the integrated soft flux in 2006 was clearly larger than in 2021, and would side-step the problem of the accretion rate apparently being approximately consistent over the 21 and 15.5~yr inter-eruption intervals (as discussed in Section~\ref{quiescence}). As noted in that section, there is some indication that during the final five years before an eruption recurs, there is a slight brightening in the optical, possibly indicative of a change occurring in the accretion disc. Alternatively, it may suggest that energy from the incipient nuclear burning is reaching the surface some years before the actual thermonuclear runaway. Certainly this five year brightening interval is longer than the expected duration of a DN outburst, which is in the range of 50--550~day in the Bollimpalli models.

Closer in time to each eruption than this, \cite{adamakis11} identify a pre-outburst signal in the optical light-curves of RS~Oph, up to $\sim$~450~d (one orbital period) before the subsequent nova, which they suggest may be caused by variable mass transfer. \cite{bollimpalli18} note that their combination nova model might explain this increase.

It is still unclear whether the RG in RS~Oph fills its Roche Lobe \citep[though it seems likely it does, e.g.,][]{brandi09,booth16}, or if accretion onto the WD occurs via stellar wind capture -- or a combination of the two. While wind accretion does not {\em require} that an accretion disc exists, it also does not preclude one, and it is probable that there is a disc in the RS~Oph system \citep[see discussion by][]{wynn08}. Such a structure would naturally be required for any form of disc instability to occur.
Following the 2006 eruption, \cite{worters07} concluded that accretion was likely re-established between days 117 and 241, a minimum of several weeks after the intervals we are comparing in the current work, based on the reappearance of optical flickering. However, as \cite{sok08} suggest, the earlier cessation of flickering noted by \cite{zamanov06} could just have been a sign of a change in the inner region of the accretion disc, rather than its complete destruction. \cite{booth16} also find that the disc survives the nova in most of their simulations.
\cite{somero17} comment that their high-resolution optical spectroscopy of RS~Oph obtained 2--3~yr after the 2006 eruption did not show the double-peaked emission line profiles expected from an accretion disc, but note that this may be because the inclination of the system is too low. 

A difficulty with the combination nova model, however, is the fact that the optical light-curves are identical each time. 
The ejected mass (and hence the optical emission) following a nova explosion is very unlikely to be constant between eruptions spontaneously triggered in this manner.

\subsection{QPO persistence}

The fact that the QPO remains around 35~s during the distinctly different SSS phases following the two eruptions appears to be more consistent with a rotation explanation, rather than a pulsation mechanism \citep[see][for discussions about QPO provenance]{andy08, julo11, ness15,wolf18}. However, this begs the question of why the SSS is not emitting smoothly over the entire surface of the WD, as might typically be expected during a TNR. A possible explanation could be the presence of a magnetic field on the WD; this could manifest in a number of ways:

\begin{itemize}
\item The nuclear burning rate could be suppressed by the enhanced magnetic pressure at the poles, causing it to be lower. This is somewhat similar to the effect which leads to sunspots, whereby magnetic fields suppress the underlying convection, leading to cooler areas. Assuming the magnetic and rotational axes are misaligned, then this would lead to an observed flux modulation as the WD spins. 

\item If accretion is occurring during the SSS phase, the stream could be magnetically funnelled towards the polar regions, causing a nuclear burning hotspot which would move in and out of view as the WD rotates. For examples of SSS with probable signs of magnetic field confinement, see \cite{king02,aydi18,drake21}. V1500~Cyg was a nova outburst on a highly magnetised WD, where a probable hotspot of enhanced nuclear burning occurred \citep{stockman88}. \cite{shara82} discusses localised TNRs in more detail.
If the luminosity is (super-)Eddington, however, accretion is not expected - at least, not fully spherical accretion.
\end{itemize}

In both of these possible cases, the majority of the SSS emission would come from the TNR spread over the full WD surface, with only small areas affected by magnetic fields leading to the detected modulation \citep[see discussion in ][]{aydi18}. While we are unaware of independent measurements of a magnetic field in RS~Oph, coherent X-ray modulation is usually taken as sufficient evidence for this in accreting systems, and one is assumed in the model by \cite{bollimpalli18}, as mentioned in Section~\ref{comb}.

A simple rotation effect would imply a fixed period, rather than one which varies, albeit slightly, as shown by the coherence measurements (although a low coherence can also indicate that the phase of the oscillation is changing, leading to a broader QPO signal, while the period remains fixed. This was seen during the 2006 eruption of RS~Oph; Beardmore et al. in prep.). Cal~83 \citep{crampton87, vandenheuvel92} is a persistent SSS: a WD accreting at a sufficiently high rate such that (quasi-)steady hydrogen burning occurs \citep{kahabka97}. This source also shows a QPO ($\sim$~67~s), and \cite{odendaal14} suggested that the X-rays might originate from an extended atmosphere, only loosely coupled to the WD, so allowing differential, non-synchronous rotation \citep[see also][]{warner02}.

\subsection{Post-SSS emission}

\begin{figure}
\begin{center}
  \includegraphics[clip, angle=-90, width=8cm]{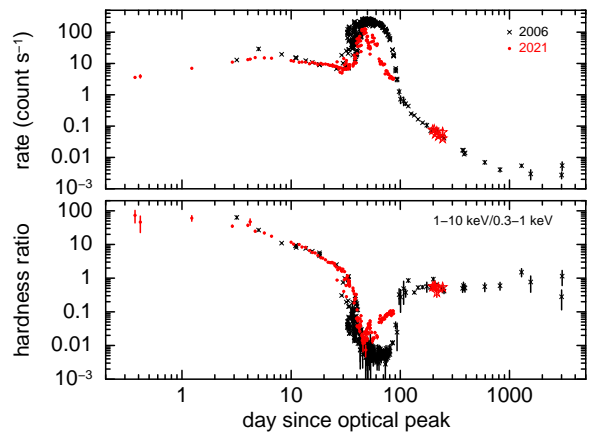}
  \caption{0.3--10~keV light-curves from 2006 and 2021, including the post-SSS data. The red stars highlight the post-Sun-constraint observations in 2022 Feb.-Apr.}
\label{late}
\end{center}
\end{figure}

After several months in Sun constraint, late-time observations of RS~Oph by {\em Swift} were obtained in 2022 Feb.--Apr. Fig.~\ref{late} shows that, at this time, the 2021 and 2006 eruption X-ray light-curves and hardness ratios are in good agreement. The X-ray spectrum now shows little evidence for any residual SSS emission, so the underlying shock continuum continues to be the same almost 200 days post eruption. As first suggested by \cite{krautter96} for V1974~Cyg, and later discussed by \cite{julo11} for RS~Oph in 2006, the X-rays seen during the decline from SSS peak may come from energy radiated as the remaining extended WD atmosphere undergoes gravitational contraction, and relaxes back onto the WD surface. 

\section{Summary}

RS~Oph re-erupted in 2021, 15.5~yr after the previous nova event, and was monitored in detail by the {\it Neil Gehrels Swift Observatory} both times, the first Galactic recurrent nova for which this has been possible. Lessons learnt from the first eruption led to the 2021 hard emission being monitored earlier and more frequently than in 2006, while the SSS phase was again followed closely. In this paper we have compared the X-ray emission from these outbursts, together with another observed by {\em EXOSAT} in 1985.

The overall optical light-curves were very similar in 2006 and 2021, suggesting the WD mass, ejecta mass, ejecta velocity distribution and ejecta geometry have remained the same. The harder X-ray emission also indicates the ejecta and immediate circumstellar environment parameters have not changed significantly, since the shocks were very similar both times. The consistency in the 2006 and 2021 outbursts of the [Fe\,{\sc x}] evolution is likely caused by a very similar duration, and luminosity, of nuclear burning.

The observed soft X-ray emission, however, was much brighter in 2006, with the ratio of the observed integrated soft X-ray counts being around 4--5$\times$ higher than in 2021. This is a much larger factor than the difference in the quiescent intervals, so is unlikely to be due to a simple change in accreted mass. There is no clear sign of a change in accretion rate during the inter-eruption intervals, either. 

While a higher absorbing column in 2021 would seem to be a simple explanation for the fainter X-rays detected, the fits to our spectra do not support this, with the measured N$_{\rm H,wind}$ being very similar in both 2006 and 2021 until day 60. A notable difference to come out of the spectral fitting, however, is a smaller BB emitting area in 2021 from around two months after the eruption. It is not immediately obvious whether this smaller radius is related to the expected shrinking of the bloated WD atmosphere as nuclear burning comes to an end, since the [Fe\,{\sc x}] measurements imply the nuclear burning duration is similar for both eruptions. There is also a brief, earlier interval around day 53 in 2021, at the time when the observed soft flux decreased, where we also see a temporary drop in the effective BB radius. 

Given that the differing optical and UV line fluxes may lean towards the ionization along our line of sight to RS~Oph in 2006 being enhanced compared with 2021 (the orbital phases at the start of the two eruptions were almost 180$^{\circ}$ apart), it seems possible that our assumption of neutral absorption in the X-ray band may be an over-simplification.  

In summary, while the optical magnitude, [Fe\,{\sc x}] and hard X-ray light-curves appear very similar following both 2006 and 2021 eruptions, the observed soft X-ray emission was found to be much brighter in 2006 than in 2021 or 1985. 
We have explored some possible explanations for the difference in the
supersoft phase, but our modelling of the {\em Swift}-XRT data does not
provide any definitive conclusions. A reduced nuclear burning luminosity
could explain the fainter X-rays in 2021 compared with 2006, although the
[Fe\,{\sc x}] line emission suggests a similar luminosity both times. If indeed
the luminosity was consistent following the separate eruptions, then the
SSS X-ray emission in the line of sight must be partly blocked, though the mechanism for this is unclear.

Looking ahead to the next RS~Oph eruption, it is likely that Athena\footnote{Advanced Telescope for High Energy Astrophysics} will have launched, and should be able to obtain high resolution, high signal-to-noise X-ray spectra. In general, UV and X-ray spectra in the late phase of the outburst will help to shed more light on the evolution of the nova ejecta; the proximity to the Sun during this latest event precluded this.

\section{Data availability}

The X-ray data underlying this article are available in the {\em Swift} archive at \url{https://www.swift.ac.uk/swift\_live/} and the HEASARC Browse archive at \url{https://heasarc.gsfc.nasa.gov/cgi-bin/W3Browse/w3browse.pl}. 

\section*{ACKNOWLEDGEMENTS}
\label{ack}

We thank the {\it Swift} PI, and Science and Mission Operations Teams for their continuing support.
KLP, APB, PAE and NPMK acknowledge funding from the UK Space Agency. MJD acknowledges funding from the UK Science and Technology Facilities Council.
This work made use of data supplied by the UK Swift Science Data Centre at the University of Leicester.

\bsp	
\label{lastpage}
\end{document}